\documentclass[fleqn, usenatbib]{mnras}

\usepackage{newtxtext,newtxmath}

\usepackage[T1]{fontenc}

\DeclareRobustCommand{\VAN}[3]{#2}
\let\VANthebibliography\thebibliography
\def\thebibliography{\DeclareRobustCommand{\VAN}[3]{##3}\VANthebibliography}

\usepackage{graphicx}	
\usepackage{amsmath}	

\newcommand{\kms}{km~s$^{-1}$}
\newcommand{\vr}{V$_{r}$}

\newcommand{\teff}{$T_{\rm eff}$}
\newcommand{\msun}{M$_\odot$}
\newcommand{\lsun}{L$_\odot$}
\newcommand{\rsun}{R$_\odot$}
\newcommand{\msunyr}{M$_\odot yr^{-1}$}
\newcommand{\vsini}{$v\sin i$}
\newcommand{\vmic}{$V_{\rm mic}$}
\newcommand{\vmac}{$V_{\rm mac}$}
\newcommand{\vrad}{$V_{\rm r}$}
\newcommand{\rstar}{$R_\star$}
\newcommand{\mstar}{$M_\star$}
\newcommand{\lstar}{$L_\star$}
\newcommand{\prot}{$P_{\rm rot}$}
\newcommand{\logg}{$\log g$}

\title[Accretion process on the enigmatic binary DQ Tau]{Accretion process, magnetic fields, and apsidal motion in the pre-main sequence binary DQ Tau}

\author[K. Pouilly et al.]{
Kim Pouilly$^{1}$\thanks{E-mail: kim.pouilly@physics.uu.se},
Oleg Kochukhov$^{1}$,
Ágnes Kóspál$^{2,3,4,5}$,
Axel Hahlin$^{1}$,
Andres Carmona$^{6}$,
and Péter Ábrahám$^{2,4,5}$
\\
$^{1}$Department of Physics and Astronomy, Uppsala University, Box 516, SE-75120 Uppsala, Sweden\\
$^{2}$Konkoly Observatory, Research Centre for Astronomy and Earth Sciences, E\"{o}tv\"{o}s Lor\'{a}nd Research Network (ELKH), Konkoly-Thege Mikl\'{o}s \'{u}t 15-17, \\ 1121 Budapest, Hungary\\
$^{3}$Max Planck Institute for Astronomy, Ko\"{n}igstuhl 17, 69117 Heidelberg, Germany\\
$^{4}$ELTE E\"{o}tv\"{o}s Lora\'{n}d University, Institute of Physics, Pa\'{z}m\'{a}ny P\'{e}ter s\'{e}t\'{a}ny 1/A, 1117 Budapest, Hungary\\
$^{5}$CSFK, MTA Centre of Excellence, Konkoly-Thege Mikl\'os \'ut 15-17, 1121 Budapest, Hungary\\
$^{6}$Univ. Grenoble Alpes, CNRS, IPAG, 38000 Grenoble, France
}

\date{Accepted 2022 November 9. Received 2022 November 9; in original form 2022 September 19}

\pubyear{2022}

\begin{document}
\label{firstpage}
\pagerange{\pageref{firstpage}--\pageref{lastpage}}
\maketitle

\begin{abstract}
Classical T Tauri stars (CTTSs) are young stellar objects that accrete materials from their accretion disc influenced by their strong magnetic field. The magnetic pressure truncates the disc at a few stellar radii and forces the material to leave the disc plane and fall onto the stellar surface by following the magnetic field lines. However, this global scheme may be disturbed by the presence of a companion interacting gravitationally with the accreting component. This work is aiming to study the accretion and the magnetic field of the tight eccentric binary DQ Tau, composed of two equal-mass ($\sim$ 0.6 \msun ) CTTSs interacting at different orbital phases.
We investigated the variability of the system using a high-resolution spectroscopic and spectropolarimetric monitoring performed with ESPaDOnS at the CFHT. We provide the first ever magnetic field analysis of this system, the Zeeman-Doppler imaging revealed a stronger magnetic field for the secondary than the primary (1.2 kG and 0.5 kG, respectively), but the small-scale fields analysed through Zeeman intensification yielded similar strengths (about 2.5 kG). The magnetic field topology and strengths are compatible with the accretion processes on CTTSs. Both components of this system are accreting, with a change of the main accretor during the orbital motion. In addition, the system displays a strong enhancement of the mass accretion rate at periastron and apastron. We also discovered, for the first time in this system, the apsidal motion of the orbital ellipse.
\end{abstract}




\section{Introduction}

A large fraction of low-mass stars have been formed in multiple systems.
Nevertheless, the accretion processes, key processes of the stellar evolution, of classical T Tauri (CTTS) stars have been mostly studied on single objects \citep[e.g.][]{Bouvier07a, Alencar12, Alencar18, Donati19, Pouilly20, Pouilly21}.
The CTTSs are pre-main sequence objects, accreting material from their circumstellar discs.
The strong magnetic field of these objects exerts a magnetic pressure on the disc at a few stellar radii (i.e. the truncation or magnetospheric radius), truncating the disc by forcing the material to follow the magnetic field lines.
This forms the so-called accretion funnel flows, columns of material following the magnetic field lines at near free-fall velocity from the disk to the stellar surface.
At the stellar surface, the kinetic energy of the accretion funnel flow is dissipated in an accretion shock, heating this area and forming the so-called hotspots \citep[see reviews in][]{Bouvier07b, Hartmann16}. 
The different signatures of the magnetospheric accretion process can be studied through spectroscopic and spectropolarimetric time series obtained over several stellar rotation cycles.
The stellar surface, the accretion funnel flows, and the inner edge of the accretion disc are (almost) corotating, producing a modulation of the observables on the stellar rotation period.
The accretion funnel flow contracts the material and heats it, giving rise to the emission of spectral lines such as the Balmer series \citep{Muzerolle01}, and due to high accretion shock temperature, narrow emission component appears in some lines (i.e. accretion tracers) such as the Ca~\textsc{ii} infrared triplet (IRT) or He~\textsc{i} at 587.6 nm \citep{Beristain01}.
Those diagnostics allow to spectroscopically trace the accretion, and since magnetic field shapes the accretion process, it is important to know the magnetic characteristics of the star.
As the Zeeman splitting of spectral lines in magnetic field results in circularly polarised light, high-resolution spectropolarimetry is an ideal tool study simultaneously the stellar magnetic field topology and the accretion process.

Numerous CTTSs have been studied using spectropolarimetry, but only few studies have been targeted to investigate multiple systems (e.g. see the studies of the binaries V4046~Sgr by \citealt{Donati11} and V1878~Ori by \citealt{Lavail20}).
This work aims to characterise the magnetic topology and the accretion of DQ Tau (RA 04$^{\rm h}$ 46$^{\rm min}$ 53$^{\rm s}$.058, Dec +17$^{\circ}$ 00' 00''.14), a spectroscopic binary system composed of two CTTSs accreting material from a circumbinary disc \citep{Basri97, Mathieu97}.
DQ Tau is en extremely interesting system because the two components of this system are orbiting each other on a 15.8 d period on a very eccentric orbit ($e \sim$ 0.6).
The separation at periastron is short enough (12.5 R$_\odot$) to allow an interaction of the two magnetospheres \citep{Czekala16}, leading to radio flares and magnetic reconnection events \citep{Salter10}.

Given what is known about the magnetically driven accretion of CTTSs, and the gravitational perturbation induced by such a close orbit, an effect on the accretion of material is expected.
Indeed, \cite{Tofflemire17} proposed a "pulsed accretion" phenomenon, an enhancement of the mass accretion rate by an order of magnitude close to the periastron passage.
These interactions at periastron have also been identified using the Kepler K2 light curve, which shows strong bursting events at each periastron passage, and sometimes at apastron as well, superimposed on a sinusoidal modulation typical of a spotted surface \citep{Kospal18}. 
The latter allowed the authors to derive a precise rotation period attributed to the primary, 3.017 d, caused by large surface cool spots.
Recently, \cite{Fiorellino22} analysed the accretion of DQ Tau at 8 epochs using X-Shooter spectra acquired over 7 orbital periods.
The authors found the two components to be similar and showed that the veiling (additional continuum emitted by a hotspot) varies with the orbital phase.
Those authors also detected the pulsed accretion for this system, an found that the main accretor is changing between the A and B component.

In this work we perform an analysis of 18 high-resolution optical spectropolarimetric observations of DQ Tau, with a one-day sampling, thus covering slightly more than one orbital cycle.
This data set allowed us to derive for the first time the magnetic structure of this system and study the accretion of both components along a complete orbital cycle.
The aim of this study is to connect the accretion and the magnetic field of both components during the orbital motion where an interaction between the two magnetospheres occurs at periastron.

This paper starts with a description of the spectropolarimetric observations used in this work (Sec.~\ref{sec:obs}).
Section \ref{sec:results} presents the results of our study, including the derivation of least-squares deconvolved (LSD) profiles, radial velocities, and orbital parameters (Sec.~\ref{subsec:LSD}), stellar fundamental parameters estimated from disentangled spectra of both components (Sec.~\ref{subsec:specdis}), a study of the strongest emission lines of DQ Tau spectrum (Sec. \ref{subsec:emLines}), and a mass accretion analysis (Sec. \ref{subsec:macc}). This section ends with a study of both large- and small-scale magnetic fields of the DQ Tau components using Zeeman-Doppler imaging (ZDI, Sec.~\ref{subsec:zdi}) and Zeeman intensification analyses (Sec.~\ref{subsec:zeemanInt}).
Finally, we discuss our results in Sec. \ref{sec:discuss} and conclude this work in Sec. \ref{sec:ccl}.

\section{Observations}
\label{sec:obs}

The observations used in this work were performed using the Echelle SpectroPolarimetric Device for the Observation of Stars \citep[ESPaDOnS,][]{Donati03} mounted at the Canada-France-Hawaii telescope (CFHT).
They consist of 18 observations, acquired between 2020 November 21 and 2020 December 8, covering the 300--1000 nm wavelength range with a resolving power of 68,000 in spectropolarimetric mode.
Each observation is composed of four sub-exposures taken in different polarimeter's configurations, which are then combined in order to obtain the intensity (Stokes I), the circularly polarised (Stokes V), and the null polarisation spectra.
The complete set of exposures was reduced using the Libre-ESpRIT package \citep{Donati97}.
The signal-to-noise ratio (S/N) of the intensity spectra is ranging between 88 and 140, while the polarised one is ranging from 65 to 118, except for the observation at HJD 2~459~177.9018 (2020 November 24) were it goes down to 54 and 18 for Stokes I and V, respectively, due to bad weather. 
The log of observations is presented in Table \ref{tab:logObs}.

\begin{table}
    \centering
        \caption{Log of ESPaDOnS observations of DQ Tau. The columns are listing the calendar and heliocentric Julian dates of observations, the S/N for the spectral pixel at the order 31 (731 nm) for the Stokes I and V spectra, the effective S/N of the LSD Stokes V profiles (see Sec.~\ref{subsec:LSD}), and the orbital phases computed from the orbital elements derived in this work (Sect.~\ref{subsec:LSD}).}
    \begin{tabular}{l l l l l l}
        \hline
        \hline
        Date & HJD & S/N$_{\rm I}$ & S/N$_{\rm V}$ & S/N$_{\rm LSD}$ & $\phi_{\rm orb}$\\ 
        (2020) & ($-$2~450~000 d) & & & & \\
        \hline
        21 Nov & 9174.8755 & 131 & 118 & 5430 & 0.07 \\
        24 Nov & 9177.9018 & 54 & 18 & 830 & 0.26 \\
        25 Nov & 9178.8047 & 88 & 65 & 2734 & 0.32 \\
        26 Nov & 9179.8423 & 107 & 93 & 4239 & 0.39 \\
        27 Nov & 9180.7957 & 122 & 105 & 4890 & 0.45 \\
        28 Nov & 9181.8613 & 131 & 118 & 5430 & 0.51 \\
        29 Nov & 9183.0852 & 119 & 101 & 4656 & 0.59 \\
        30 Nov & 9183.9523 & 111 & 99 & 4773 & 0.65 \\
        01 Dec & 9185.0181 & 140 & 127 & 5881 & 0.71 \\
        02 Dec & 9185.8267 & 128 & 114 & 5250 & 0.76 \\
        03 Dec & 9186.9894 & 129 & 113 & 5431 & 0.84 \\
        03 Dec & 9187.0596 & 131 & 116 & 5391 & 0.84 \\
        04 Dec & 9187.9031 & 132 & 114 & 5491 & 0.90 \\
        05 Dec & 9188.9175 & 90 & 77 & 3227 & 0.96 \\
        06 Dec & 9190.0344 & 130 & 115 & 5274 & 0.03 \\
        07 Dec & 9190.9315 & 133 & 116 & 5518 & 0.09 \\
        08 Dec & 9191.8353 & 133 & 118 & 5582 & 0.15 \\
        08 Dec & 9192.0633 & 120 & 108 & 4941 & 0.16 \\

    \hline
    \end{tabular}
    \label{tab:logObs}
\end{table}

Each reduced spectrum was then normalised using a polynomial fit over semi-automatic continuum points selection with the algorithm presented in \cite{Folsom16}, ensuring consistent normalisation of overlapping parts of spectral orders and producing a flat continuum.

\section{Results}
\label{sec:results}

In this section, we describe the results obtained from the ESPaDOnS spectra described in Sec. \ref{sec:obs}. 
This consist of characterising the spectral and spectropolarimetric variations of DQ Tau over the observation time range.

\subsection{Least-squares deconvolution and radial velocity measurements}
\label{subsec:LSD}

\begin{figure}
    \centering
    \includegraphics[width=.45\textwidth]{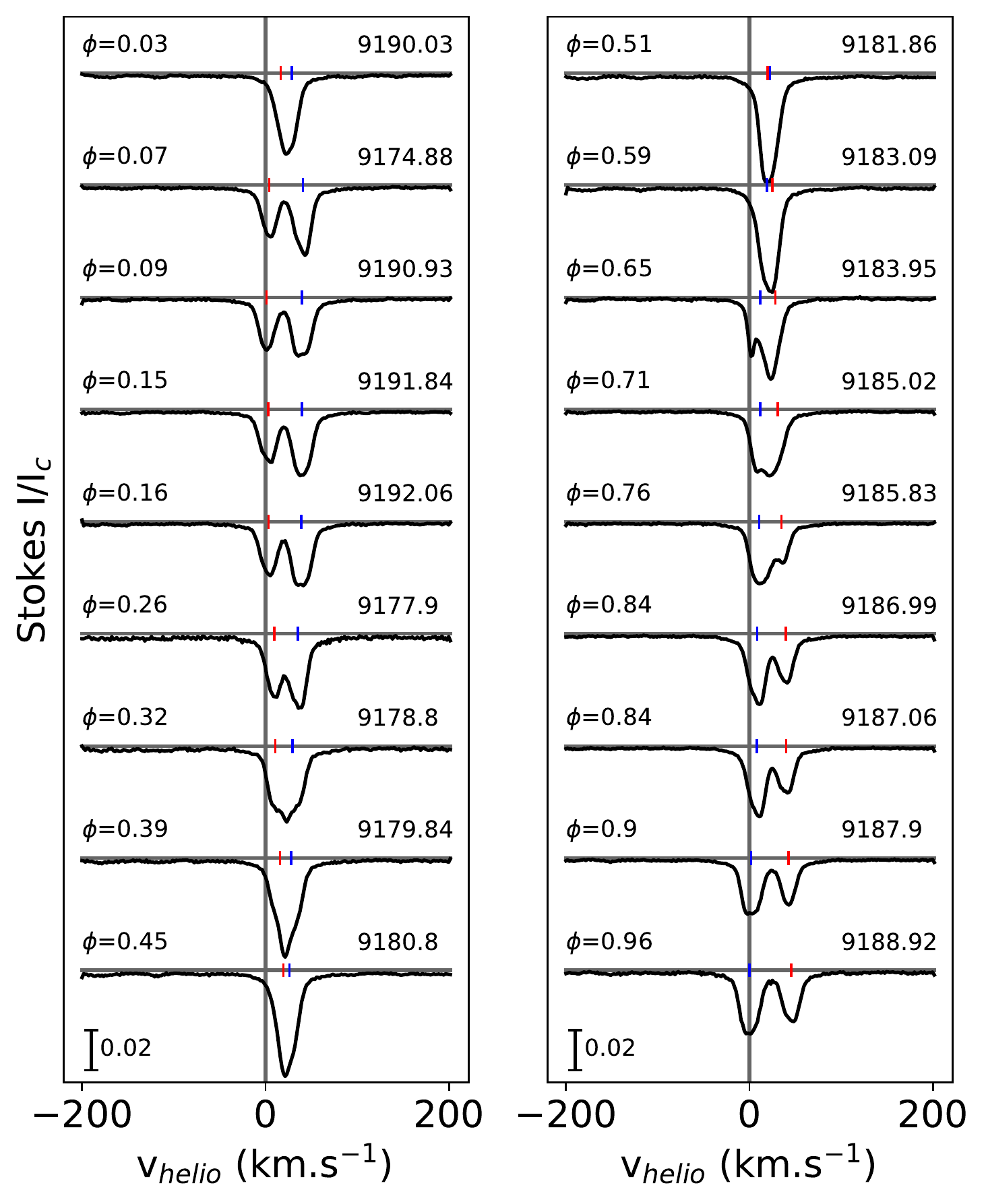}
    \caption{LSD Stokes I profiles of DQ Tau system. The orbital phase of each observation is indicated at the top left of each profile, and the HJD (-2~450~000 d) at the top right. The vertical line indicates the 0 velocity. The blue and red ticks indicate the radial velocity of the A and B component, respectively.}
    \label{fig:stokesILSD}
\end{figure}

\begin{figure}
    \centering
    \includegraphics[width=.45\textwidth]{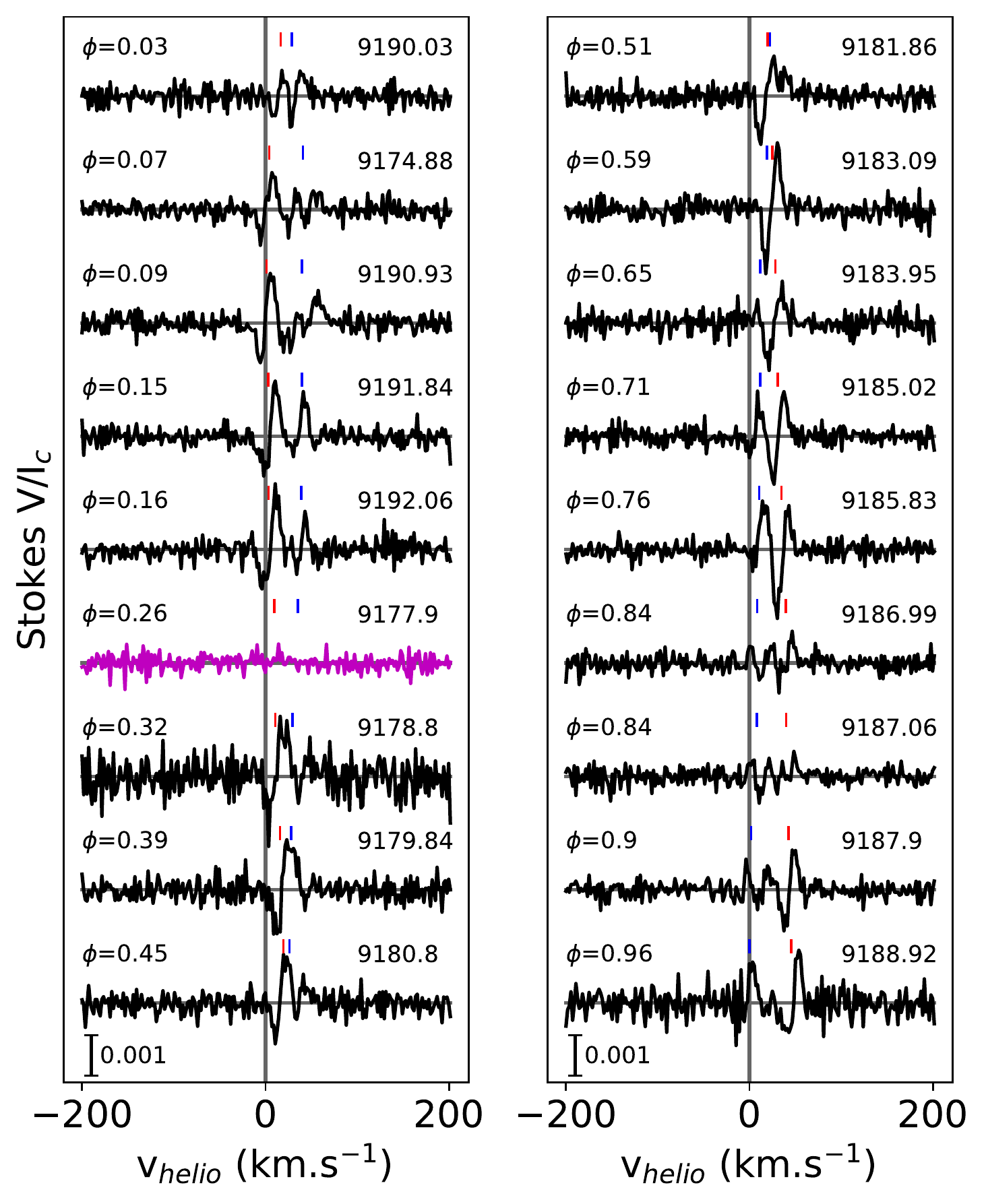}
    \caption{Same as Fig. \ref{fig:stokesILSD} for LSD Stokes V profiles. As the observation at $\phi$=0.26 \textit{(magenta)} exhibits a poorer S/N, we divided this profile (only) by 10 for more visibility.}
    \label{fig:stokesVLSD}
\end{figure}

We started this analysis by computing the least-squares deconvolution \citep[LSD,][]{Donati97} profiles for the ESPaDOnS spectra, and more specifically the Stokes I (unpolarized) and Stokes V (circularly polarized) profiles.
This method allows one to drastically increase the S/N of the Zeeman signatures by computing a kind of weighted mean of those signatures over as many photospheric lines as possible, meaning about 12~000 lines in this work.
We used a mean wavelength of 520 nm, an intrinsic line depth of 0.2 and a Landé factor of 1.2 for normalisation of LSD weights.
The derivation used a line list that we extracted from the VALD database \citep{Ryabchikova15}, covering the ESPaDOnS wavelength range.
We then removed the regions contaminated by emission lines or telluric absorption.
The resulting Stokes I and V profiles are shown in Fig. \ref{fig:stokesILSD} and Fig. \ref{fig:stokesVLSD}, respectively.
They exhibit an effective S/N ranging from 880 to 1067 for the Stokes I profiles, and from 830 to 5890 for the Stokes V profiles.
The LSD Stokes I profiles illustrate the motion of the two components in velocity space, with the deeper component (i.e., the primary) and the shallower one (i.e., the secondary) getting closer, merging, and move away during the observation time range, as expected from the orbital period of 15.8 days.
The LSD Stokes V profiles exhibit detected signatures at all observations, except for the one at HJD 2~459~177.9018, as expected from the low S/N of this observation (see Table \ref{tab:logObs}).
Those signatures are significant for both the primary and the secondary, suggesting a strong large-scale fields for both components of the system.

\begin{figure}
    \centering
    \includegraphics[width=.49\textwidth]{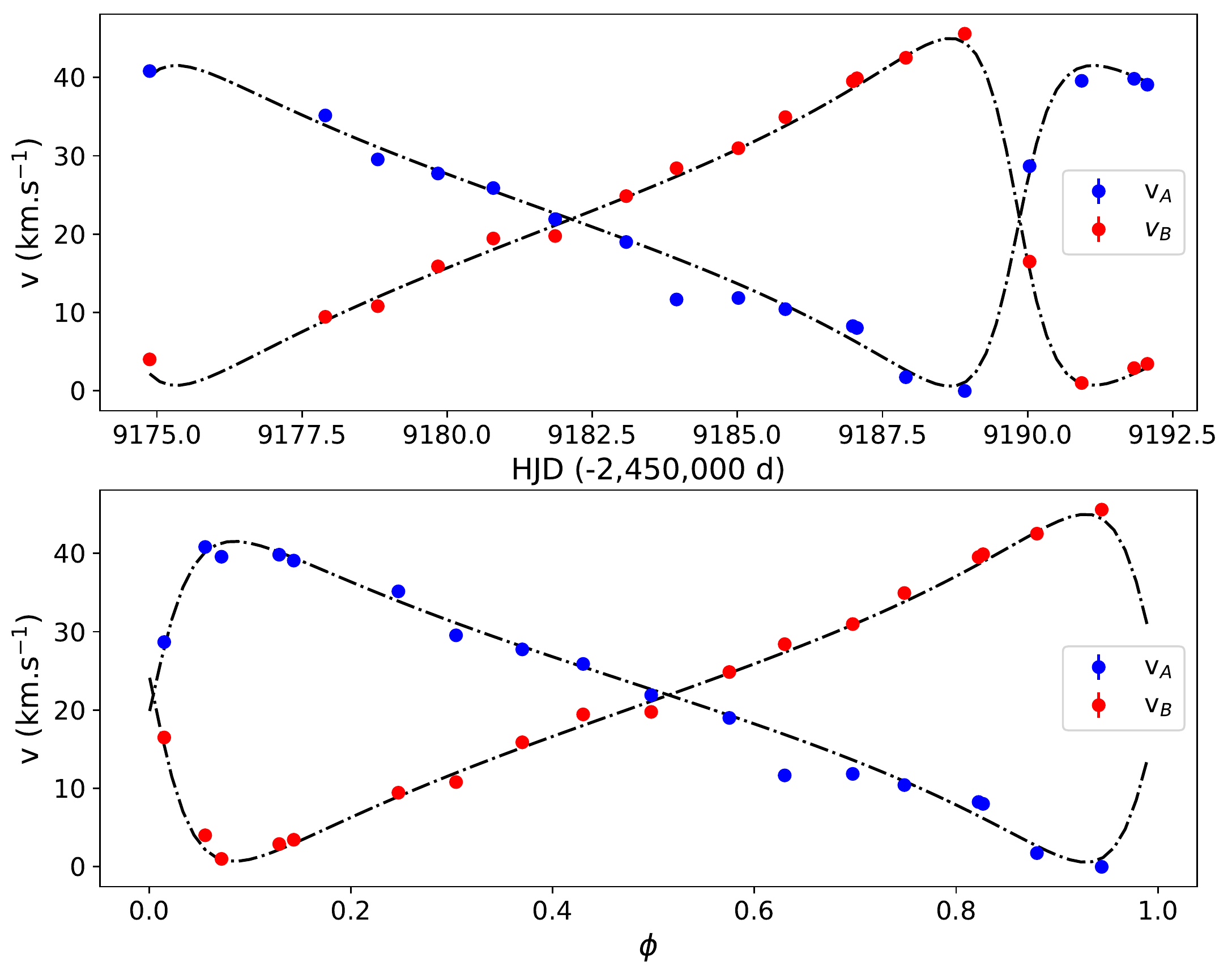}
    \caption{Radial velocities of the primary (A, in blue) and secondary (B, in red) component of the DQ Tau system, determined from the LSD Stokes I profiles, as a function of HJD \textit{(top)}, and the orbital phase \textit{(bottom)}. The error bars are provided but smaller than the symbol size. The black dash-dotted curve is the orbital solution derived in this work, with the parameters provided in Table \ref{tab:orbElements}.}
    \label{fig:vradLSD}
\end{figure}

Then we disentangled the two components of the LSD Stokes I profiles by using an iterative scheme based on the method by \cite{Gonzalez06}.
This method allows both disentangling of the profiles, producing a mean Stokes I profile for each component, and derivation of a precise estimation of the radial velocity of each component.
Initially, our algorithm assumed each component to be a Voigt profile, shifted by an initial guess of the radial velocity of each component.
For this first guess, we used the orbital solution provided by \cite{Czekala16}.
The whole profile of each observation was then shifted by the velocity of the secondary, and corrected from the primary's contribution.
The radial velocity of the B component is thus adjusted at this moment by fitting a Voigt profile.
All profiles corrected are then averaged, producing a mean LSD Stokes I profile of the secondary.
Then we performed this step for the A component, and repeated those two steps until a convergence criterion was reached.
As a criterion we set the convergence when the radial velocity correction is lower than 10$^{-2}$ \kms\ for both components.
The velocities obtained for both components are summarized in Table. \ref{tab:vrad} and illustrated in Fig. \ref{fig:vradLSD}.

\begin{table}
    \centering
        \caption{Radial velocities of DQ Tau's A-B pair and their uncertainties computed from the LSD Stokes I disentangling procedure.}
    \begin{tabular}{l | l l | l l}
        \hline
        HJD & \vrad(A) & $\delta$\vrad(A) & \vrad(B) & $\delta$\vrad(B) \\
        ($-$2~450~000 d) & \kms & \kms & \kms & \kms \\
        \hline
        9174.8755 & 40.82 & 0.21 & 4.00 & 0.22 \\
        9177.9018 & 35.15 & 0.34 & 9.44 & 0.33 \\
        9178.8047 & 29.54 & 0.23 & 10.80 & 0.25 \\
        9179.8423 & 27.74 & 0.26 & 15.88 & 0.23 \\
        9180.7957 & 25.88 & 0.22 & 19.44 & 0.19 \\
        9181.8613 & 21.91 & 0.24 & 19.75 & 0.21 \\
        9183.0852 & 18.99 & 0.27 & 24.85 & 0.23 \\
        9183.9523 & 11.65 & 0.44 & 28.41 & 0.44 \\
        9185.0181 & 11.84 & 0.19 & 30.97 & 0.18 \\
        9185.8267 & 10.42 & 0.35 & 34.93 & 0.31 \\
        9186.9894 & 8.25 & 0.21 & 39.53 & 0.20 \\
        9187.0596 & 8.01 & 0.21 & 39.89 & 0.21 \\
        9187.9031 & 1.73 & 0.20 & 42.51 & 0.19 \\
        9188.9175 & $-0.03$ & 0.23 & 45.58 & 0.20 \\
        9190.0344 & 28.68 & 0.41 & 16.49 & 0.48 \\
        9190.9315 & 39.57 & 0.18 & 0.99 & 0.15 \\
        9191.8353 & 39.83 & 0.20 & 2.89 & 0.21 \\
        9192.0633 & 39.08 & 0.21 & 3.43 & 0.18  \\

    \hline
    \end{tabular}
    \label{tab:vrad}
\end{table}

These radial velocities allowed us to derive the orbital parameters of DQ Tau and to check their consistency with the results of \cite{Czekala16}.
We thus fitted an orbital solution using a Levenberg--Marquart algorithm (LMA) for the two components separately, letting the following parameters varying freely: the orbital period $P_{\rm orb}$, the systemic velocity $\gamma$, the semi-amplitude $K$, the eccentricity $e$, the argument at periastron $\omega$, and the time at periastron $T_{\rm peri}$.
Then we averaged the parameters which are supposed to be the same for both components ($P_{\rm orb}$, $\gamma$, $e$, $\omega$, and $T_{\rm peri}$) to perform a new fit letting only the semi-amplitudes of both components as free parameters.
Those results, as well as the orbital parameters obtained by \cite{Czekala16}, are summarised in Table \ref{tab:orbElements} and the corresponding orbital radial velocity curves are presented in Fig.~\ref{fig:vradLSD}.

\begin{table}
    \centering
    \caption{Orbital elements of the DQ Tau system. $P_{\rm orb}$ is the orbital period, $\gamma$ is the systemic velocity, $K_1$ (and $K_2$) is the semi-amplitude of the A (and B) component, $e$ is the eccentricity, $\omega$ is the argument at periastron, and $T_{\rm peri}$ is the time at periastron. The estimations of the orbital inclination $i_{\rm orb}$, the distance at periastrion $d_{\rm peri}$ and at apastron $d_{\rm apa}$ computed from the orbital elements and the stellar atmospheric parameters (Sec. \ref{subsec:stellParam}) are also provided.}
    \begin{tabular}{l c c}
        \hline
        Parameter & \cite{Czekala16} & This work \\
        \hline
        $P_{\rm orb}$ (days) & 15.80158 $\pm$ 0.00066 & 15.80 $\pm$ 0.01 \\
        $\gamma$ (\kms) & 24.52 $\pm$ 0.33 & 21.9 $\pm$ 0.4 \\
        $K_1$ (\kms) & 20.28 $\pm$ 0.71 & 20.5 $\pm$ 0.6 \\
        $K_2$ (\kms) & 21.66 $\pm$ 0.60 & 22.2 $\pm$ 0.4 \\
        $e$ &  0.568 $\pm$ 0.013 & 0.58 $\pm$ 0.10 \\
        $\omega$ ($^{\circ}$) & 231.9 $\pm$ 1.8 & 266.0 $\pm$ 3.6 \\
        $T_{\rm peri}$ (HJD - 2~400~000) & 47433.507 $\pm$ 0.094 & 59173.73 $\pm$ 0.10 \\
        $i_{\rm orb}$ ($^{\circ}$) & - & 26 \\
        $d_{\rm peri}$ (AU) & - & 0.05 \\
        $d_{\rm apa}$ (AU) & - & 0.21 \\

    \hline
    \end{tabular}
    \label{tab:orbElements}
\end{table}

\subsection{Spectral disentangling and stellar parameters}

In this section, we derive the stellar parameters of both components of DQ Tau system.
This started with disentangling the spectra of the two components.

\subsubsection{Spectral disentangling}
\label{subsec:specdis}

In order to derive the stellar parameters of each component of DQ Tau, we have to disentangle the two components.

We used the disentangling algorithm developed by \cite{Folsom10} from \cite{Gonzalez06}' method and used in other works such as \cite{Kochukhov18}, \cite{Lavail20}, \cite{Hahlin21}.
This algorithm assumes that the composite spectrum of a binary star is composed of two components, which stay constant during the whole time series and are only shifted in radial velocity variable from one observation to another, and given by an initial guess of the orbital solution.
In the case of DQ Tau, we kept the one given by \cite{Czekala16}.
This iterative procedure allows to retrieve the mean photospheric spectrum of each component.
We applied this procedure to 11 wavelength windows, ranging from 489 to 755 nm and about 5 to 10 nm-wide each. 

As the original spectra were normalised, the continuum of the to components were diluted.
We thus rescaled the disentangled spectra of the components A and B to correct for the continuum dilution, not included in the algorithm, using the following formula \citep{Folsom08}:
\begin{align}
    S^{\rm 0}_{\rm A} \ &= \ \left( S_{\rm A} - \frac{1}{1+r_{\rm L}} \right) \ \times \ \left( 1 + \frac{1}{r_{\rm L}} \right)
    \label{eq:dilA}
    \\
    S^{\rm 0}_{\rm B} \ &= \ \left( S_{\rm B} - \frac{r_{\rm L}}{1+r_{\rm L}} \right) \ \times \ \left( 1 + r_{\rm L} \right),
    \label{eq:dilB}
\end{align}

\noindent where $S^{\rm 0}_{\rm A}$ and $S_{\rm A}$ ($S^{\rm 0}_{\rm B}$ and $S_{\rm B}$) are the corrected and diluted spectra of A (B) component, respectively, and $r_{\rm L}=L_{\rm A}/L_{\rm B}$ is the luminosity ratio of the two components. 
Given the very close properties of the components, we assumed $r_{\rm L}$ = 1, which is in agreement with the stellar luminosities given by \cite{Tofflemire17}.

\subsubsection{Stellar parameters}
\label{subsec:stellParam}

We used the disentangled spectra produced in Sec. \ref{subsec:specdis} to derive the atmospheric parameters of the DQ Tau system, meaning the effective temperature \teff, the projected rotational velocity \vsini, and the microturbulent velocity \vmic.
We fitted synthetic spectra for each wavelength window using the \texttt{ZEEMAN} code \citep{Landstreet88, Wade01, Folsom12}, which computes synthetic spectra to reproduce the observations through a LMA using MARCS atmosphere models \citep{Gustafsson08}, VALD line list \citep{Ryabchikova15}, and the same oscillator strength correction as used by \cite{Folsom16}.
We assumed a macroturbulent velocity \vmac\ = 2 \kms, a surface gravity \logg\ = 4 \citep{Tofflemire17} and a solar metallicity, wich are typical values for low-mass TTSs \citep{Padgett96, James06, Santos08, Taguchi09, Dorazi11}.
In order the reject the windows where the fitting procedure failed, we excluded the values larger than 2-sigma from the mean, before averaging the results and using the standard deviation as uncertainty.
The result of this fit over the 744 to 754 nm wavelength range is presented in Fig.~\ref{fig:ZEEMANfit}.
This yielded \teff\ = 4007 $\pm$ 53 K, \vsini\ = 15.4 $\pm$ 0.6 \kms, and \vmic\ = 1.2 $\pm$ 0.2 \kms\ for the A component. The B component exhibits very close values: \teff\ = 4050 $\pm$ 48 K, \vsini\ = 13.1 $\pm$ 0.6 \kms, and \vmic\ = 1.2 $\pm$ 0.2 \kms.
Those effective temperatures are consistent with the \teff\ = 4000 $\pm$ 300 K obtained by \cite{Czekala16}, and correspond to a K7--K8 spectral type following the conversion table of \cite{Pecaut13}, which is close to the M0 type derived in the literature from comparisons with empirical templates \citep{Herczeg14, Tofflemire17, Fiorellino22}.
The \vsini\ we derived are consistent with the values found by \cite{Nguyen12}, within 1$\sigma$ for the A component and 2$\sigma$ for the B component.

Because the DQ Tau system is composed of accreting TTSs, we considered the effect of the veiling which fills the photospheric lines and induces a wrong estimation of the stellar parameters if it is ignored.
The so-called continuum veiling arises from an additional continuum flux emitted by the hotspot.
In order to measure it, we choose a non-accreting photospheric template of about the same effective temperature observed with ESPaDOnS: V819 Tau.
This object is a WTTS with a K7 spectral type, \vr\ = 16.6 \kms\ and \vsini\ = 9.5 \kms\ \citep{Donati15}.
We rotationally broadened the template using the Gray function \citep{Gray73} and fitted the template in the wavelength windows using LMA by adding the veiling using the following formula:
\begin{equation}
    I = \frac{I_0 + r}{1 + r},
\end{equation}

\noindent where I is the veiled spectrum, $I_0$ the spectrum without veiling and $r$ is the continuum veiling.
However, \cite{Rei18} found an additional line-dependent component of the veiling, particularly present in the strongest photospheric lines. 
In order to measure only the continuum veiling, we thus performed the derivation using lines with an equivalent width (EW) between 0.001 nm and 0.01 nm, following the authors' prescriptions.
This yields $r$(A) = 0.05 $\pm$ 0.07 and $r$(B) = 0.5 $\pm$ 0.2 for A and B components, respectively.
These values indicate that the B component was accreting during our observations, and explain the weaker LSD Stokes I profiles of the B component, even assuming $r_{\rm L}$ = 1.
\cite{Fiorellino22} found (unresolved) veiling values for the DQ Tau system of $<$0.5 for the orbital phases 0.0--0.8 and $>$0.5 for the orbital phases 0.8--1.0, meaning close to periastron. 
This suggests that even though some of our spectra were taken close to periastron, the DQ Tau components were not in a very actively accreting phase during our observations, or at least less than during the observations used by \cite{Fiorellino22}.

Then we estimated the primary's luminosity using the system's magnitude in J-band $m_{\rm J}$ = $9.4197 \pm 0.0031$, computed by averaging the results of \cite{Muzerolle19} outside the periastron passage, $r_{\rm L}$ = 1, the interstellar extinction computed by \cite{Fiorellino22}, $A_{\rm V}$= $1.72 \pm 0.26$ combined with the interstellar reddening law of \cite{Cardelli89} and the Gaia EDR3 parallax, $\pi$ = $5.118 \pm 0.018$ mas \citep{Gaia21}.
This yielded $L$ = $0.935 \pm 0.081$ L$_{\odot}$, and thus $R$ = $2.011 \pm 0.086$ R$_{\odot}$ using the Stefan-Boltzmann law.
We derived the inclination angle $i$ = $27 \pm 4^{\circ}$ by combining the radius, the rotation period \prot\ = $3.03 \pm 0.04$ d from \cite{Kospal18} and the \vsini.

Finally, we placed the star in an Hertzsprung-Russel diagram (HRD) and fitted its position using a grid of CESTAM evolutionary models \citep{Marques13, Villebrun19} to obtain its mass, age, and internal structure. 
From those models, the primary is a 0.587$^{+0.052}_{-0.046}$ \msun\ fully convective star with an age of about 1 Myr.
We did the same procedure using \cite{Baraffe98} models, finding consistent results, but the parameters of DQ Tau was better recovered by the CESTAM grid, which is more convenient to performed the bi-linear interpolation needed to derive those parameters.

We derived the same parameters for the companion and obtained $R$ =  1.968 $\pm$ 0.084 \rsun, $M$ = $0.628^{+0.050}_{-0.046}$ \msun, a 1 Myr age and a fully convective interior as well.
As the rotation period is still unknown for the B component, we did not derive the corresponding inclination angle.
The whole set of parameters for each component is summarised in Table \ref{tab:param}.

The masses and the inclination angle we obtained are consistent with the results of \cite{Fiorellino22}, based on the work of \cite{Czekala16} and corrected using Gaia EDR3 distance.

\begin{table}
    \centering
    \caption{DQ Tau stellar parameters for the component A and B. The rows give the effective temperature, the rotational and the microscopic velocity, the stellar luminosity, radius, mass, age, the radius and mass of the radiative core, the rotation period, and the stellar inclination.}
    \begin{tabular}{l l l}
        \hline
        Parameter & A & B \\
        \hline
        \teff & 4007 $\pm$ 53 K & 4050 $\pm$ 48 K \\
        \vsini & 15.4 $\pm$ 0.6 \kms & 13.1 $\pm$ 0.6 \kms \\
        \vmic & 1.2 $\pm$ 0.2 \kms & 1.2 $\pm$ 0.2 \kms \\
        \lstar & 0.935 $\pm$ 0.081 \lsun & 0.935 $\pm$ 0.081 \lsun $^{\left(1\right)}$ \\
        \rstar & 2.011 $\pm$ 0.086 \rsun & 1.968 $\pm$ 0.084 \rsun $^{\left(1\right)}$ \\
        \mstar & 0.587 $^{+0.052}_{-0.046}$ \msun & 0.628 $^{+0.050}_{-0.046}$ \msun $^{\left(1\right)}$ \\
        Age & 1.1 $\pm$ 0.4 Myr & 1.2 $\pm$ 0.2 Myr $^{\left(1\right)}$ \\
        \prot & 3.03 $\pm$ 0.04 d $^{\left(2\right)}$ & - \\
        $i$ & 27 $\pm$ 4$^{\circ}$ & - \\

    \hline
    \end{tabular}
    \begin{flushleft}
    Notes: (1) Assuming a luminosity ratio of one between the two components. (2) Taken from \cite{Kospal18}.
    \end{flushleft}
    \label{tab:param}
\end{table}

\subsection{Emission lines}
\label{subsec:emLines}

In this section, we present our analysis of the three emission lines, H$\alpha$, H$\beta$, and the He~\textsc{i} 587.6~nm line.
The aim of analysing these lines is to investigate the accretion process in the DQ Tau system.
Indeed, the magnetospheric accretion of CTTSs takes place through accretion funnel flows, allowing the formation of lines such as Balmer lines or He~\textsc{i} at 587.6 nm, which are thus expected to be modulated over the stellar rotation period \citep[][for HQ Tau and V807 Tau, respectively]{Pouilly20, Pouilly21} thereby revealing the topology of the accretion flows.
We would like to precise that it is not possible to disentangle the spectrum of the primary and the secondary for the emission lines with the methodology used in Sec.~\ref{subsec:specdis}.
The lines presented in this section thus contain the contribution of the two components.

\subsubsection{Balmer lines: H$\alpha$ \& H$\beta$}
\label{subsubsec:balmer}

\begin{figure}
    \centering
    \includegraphics[width=.45\textwidth]{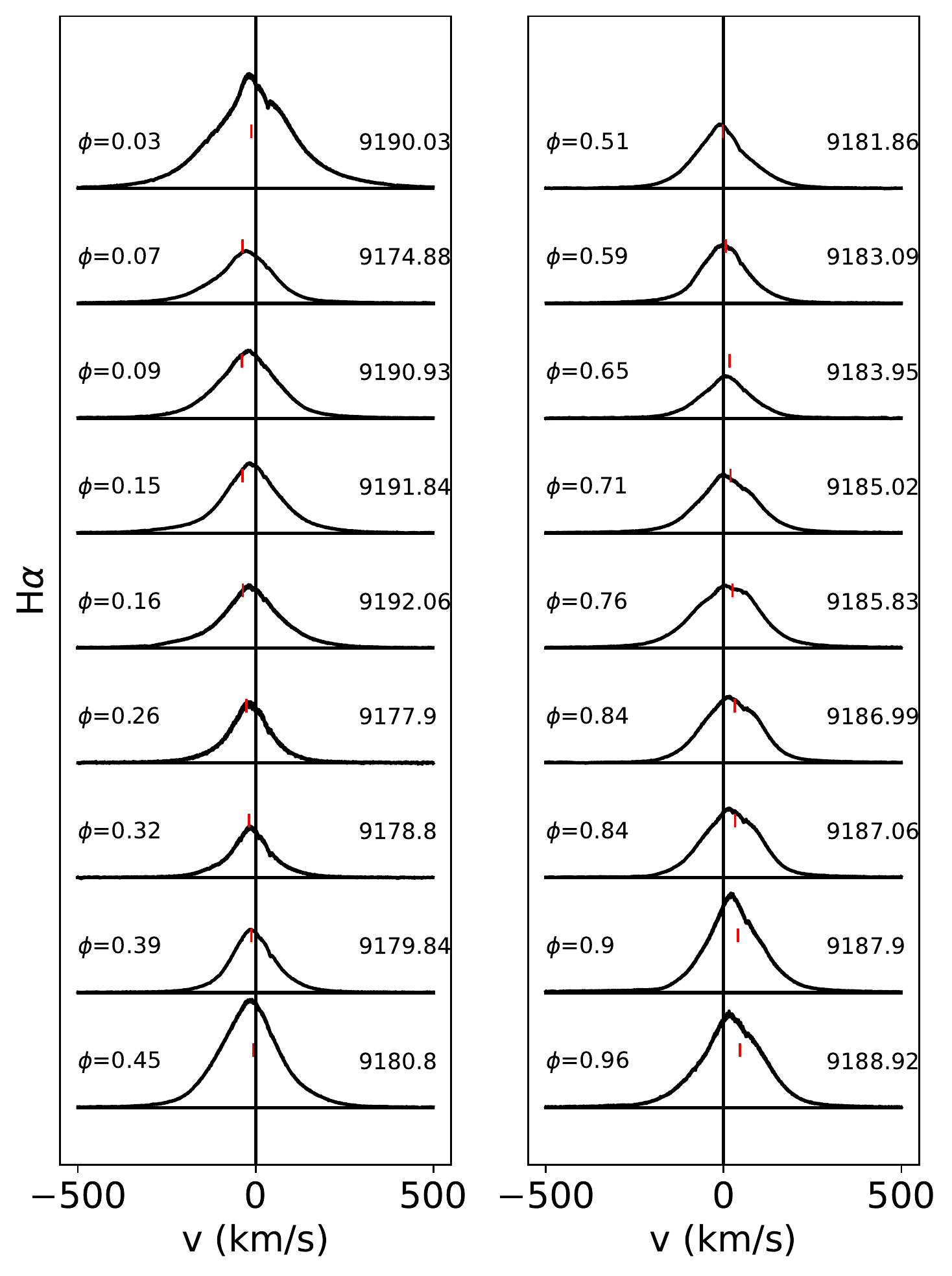}
    \caption{Residual H$\alpha$ profiles centred on the primary's radial velocity. The vertical line illustrates the 0 velocity and the red ticks indicate the radial velocity of the secondary. The orbital phase is indicated at the left of each profile and the HJD of observation is indicated on the right. }
    \label{fig:haCol}
\end{figure}

\begin{figure}
    \centering
    \includegraphics[width=.45\textwidth]{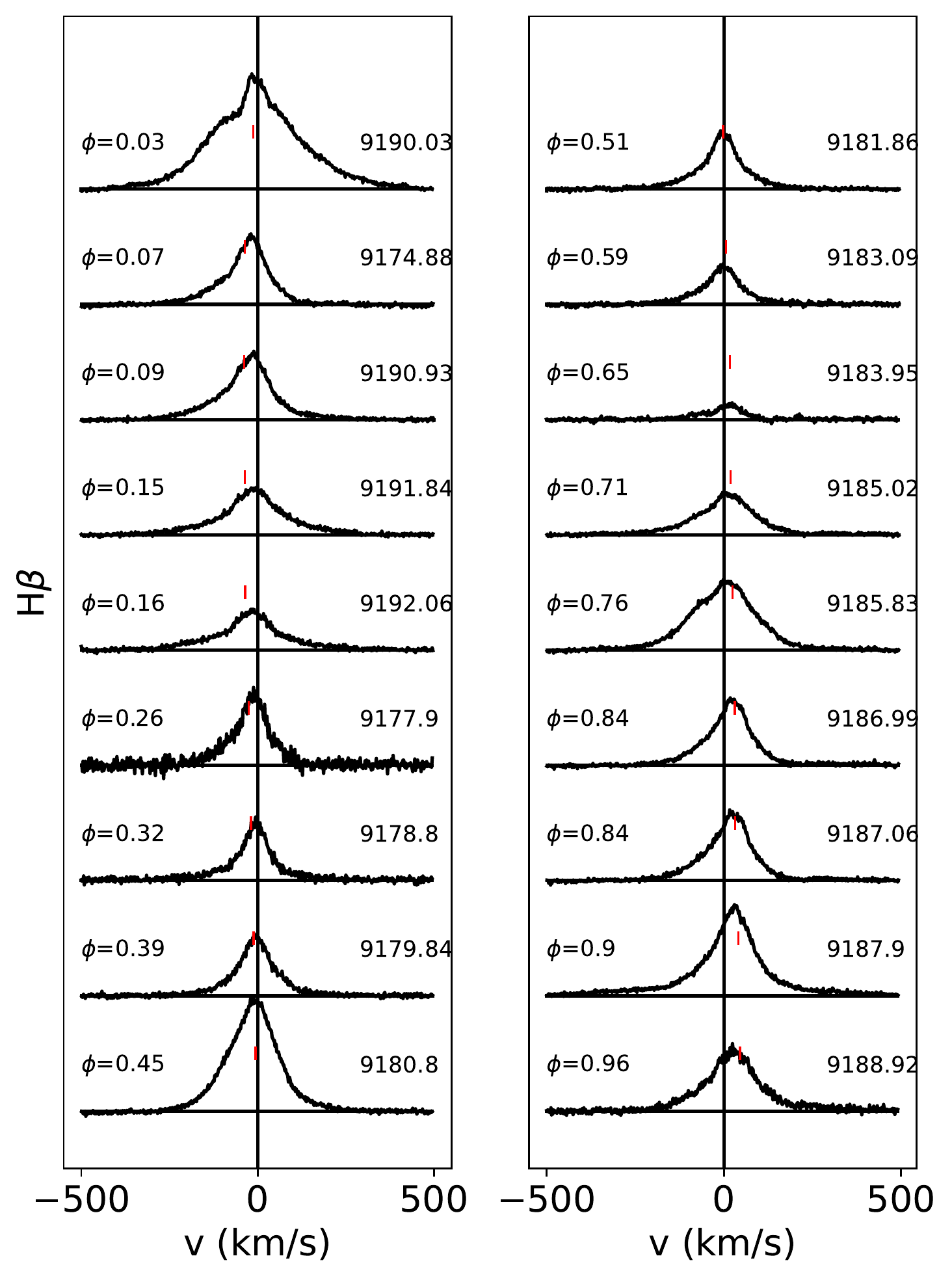}
    \caption{Same as Fig. \ref{fig:haCol} for the residual H$\beta$ profiles.}
    \label{fig:hbCol}
\end{figure}

We computed the residual profile of H$\alpha$ and H$\beta$ lines, meaning that we removed the photospheric contribution of both components by using the same photospheric template V819 Tau (see Sec. \ref{subsec:stellParam}).
As this object is a WTTS, this allowed us to remove the chomospheric contribution as well, keeping only the feature induced by the accretion.
Furthermore, we corrected the profiles for the radial velocity of the primary. 
The H$\alpha$ and H$\beta$ residual profiles are presented in Figs. \ref{fig:haCol} and \ref{fig:hbCol}, respectively.

\begin{figure}
    \centering
    \includegraphics[width=.45\textwidth]{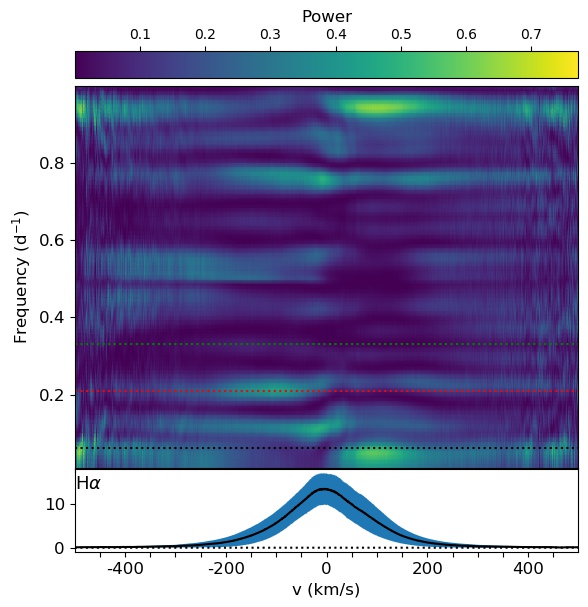}
    \caption{\textit{Top:} 2D periodogram of the residual H$\alpha$ line. The color bar goes from dark blue, for the lowest periodogram's power, to light yellow for the highest. The three horizontals dotted lines indicate the rotation period of the primary at 3.03 d \textit{(green)}, the orbital period at 15.8 d \textit{(black)} and an unexpected signal at f$\sim$0.21 d$^{-1}$ or 4.8 d \textit{(red)}. \textit{Bottom:} Mean profile \textit{(black)} and its variance \textit{(blue)}.}
    \label{fig:haPeriodo}
\end{figure}

\begin{figure}
    \centering
    \includegraphics[width=.45\textwidth]{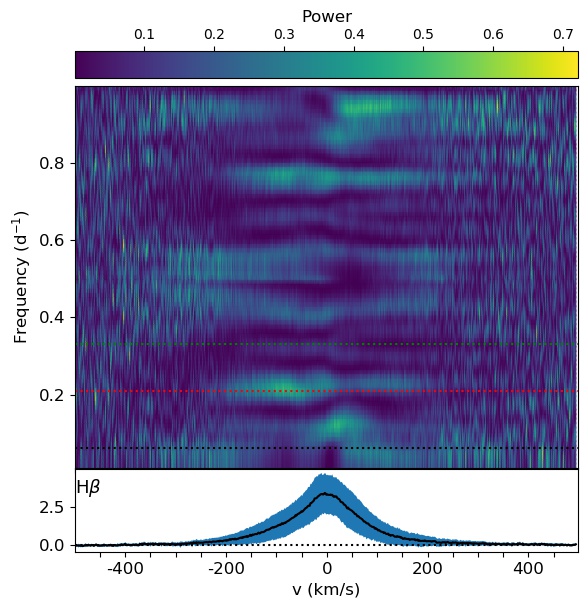}
    \caption{Same as Fig. \ref{fig:haPeriodo} for the residual H$\beta$ line.}
    \label{fig:hbPeriodo}
\end{figure}

Both line profiles exhibit a large variability in strength with a maximum reached at the orbital phases $\phi$ = 0.45 and 0.03 (HJD 2~459~180.80 and 2~459~190.03, respectively), meaning the apastron and periastron passages, respectively.

In order to investigate any periodic variation inside those Balmer lines, we computed 2D periodograms, a Lomb-Scargle periodogram \citep{Scargle82} computed in each velocity channel of the line.
The resulting 2D periodograms are presented in Figs. \ref{fig:haPeriodo} and \ref{fig:hbPeriodo} for the residual H$\alpha$ and H$\beta$ lines, respectively.
For both lines, we retrieve the orbital period at 15.8 d, mainly in the red wing, with a FAP reaching locally 10$^{-3}$.
Furthermore, the blue wing of the lines exhibits another period, around 4.8 d, with a FAP of about 10$^{-2}$, different from the primary's rotation period determined by \cite{Kospal18}.
Finally, we notice the absence of a signal at the primary's rotation period.

\subsubsection{Accretion tracer: He~\textsc{i} line}

\begin{figure}
    \centering
    \includegraphics[width=.45\textwidth]{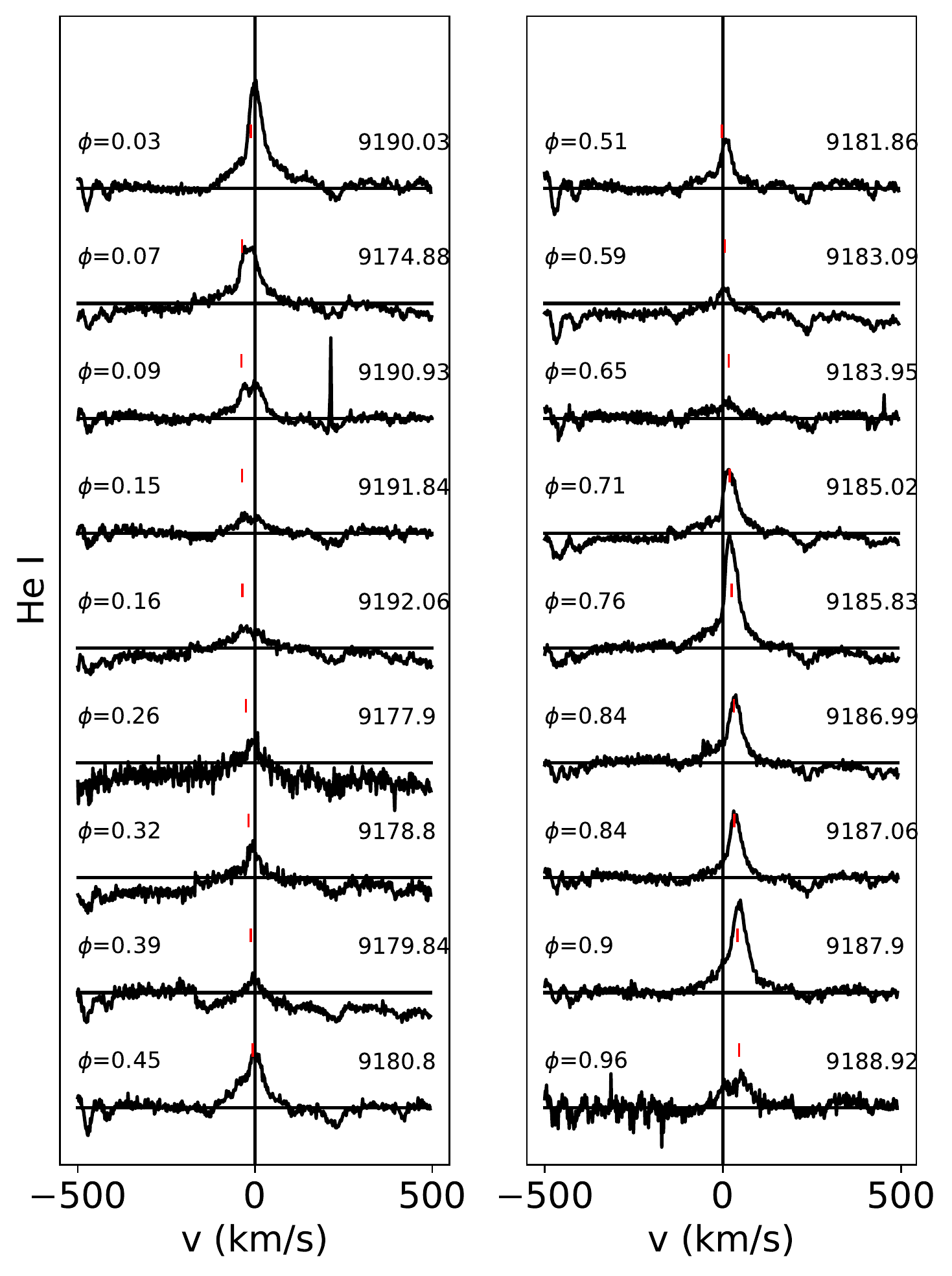}
    \caption{Same as Fig. \ref{fig:haCol} for He~\textsc{i} line at 587.6 nm. }
    \label{fig:heCol}
\end{figure}

The He~\textsc{i} line at 587.6 nm is known to be an accretion tracer. 
This is due to the fact that its narrow component (NC) is at least partly formed in the accretion post-shock region \citep{Beristain01}.
The He~\textsc{i} line of DQ Tau is presented in Fig. \ref{fig:heCol} and exhibits a maximum at phase 0.03, at the periastron passage, without any particular behavior at the apastron, unlike the Balmer lines studied in Sec. \ref{subsubsec:balmer}. 
The NC is present at all epochs, but rarely at the 0 velocity, suggesting that it is mainly emitted by the secondary.
This is consistent with the larger veiling found for the B component of the system in Sec. \ref{subsec:stellParam} and suggests that the secondary accretes more than the primary during this observing period.

\begin{figure}
    \centering
    \includegraphics[width=.45\textwidth]{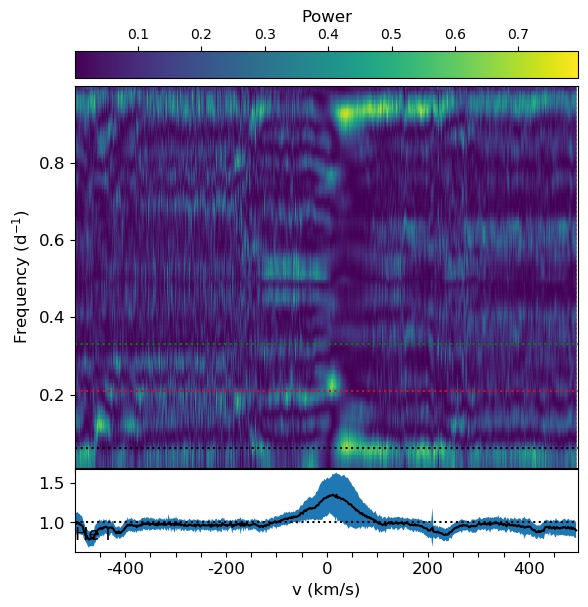}
    \caption{Same as Fig. \ref{fig:haPeriodo} for He~\textsc{i} line at 587.6 nm.}
    \label{fig:hePeriodo}
\end{figure}

The 2D periodogram of the He~\textsc{i} line is shown in Fig. \ref{fig:hePeriodo}. 
A signal at the orbital period occurs in the red wing of the line (FAP$\sim$10$^{-3}$), but the NC of the line seems to be modulated on a $\sim$5 d-period, with a FAP of 10$^{-2}$, similar to the blue wing of the Balmer lines studied in Sec. \ref{subsubsec:balmer}.
Assuming that this line is emitted by the secondary, the absence of a signal at 3 days is not surprising given that it is the rotation period of the primary, the secondary's rotation period remains unknown.

\begin{figure}
    \centering
    \includegraphics[width=.45\textwidth]{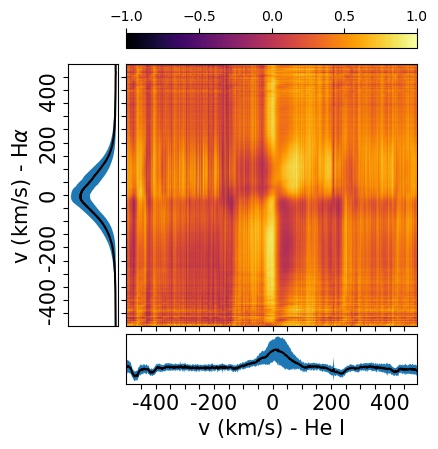}
    \caption{Correlation matrix of H$\alpha$ vs. He~\textsc{i} lines. The colormap illustrates the correlation coefficient: light yellow for a strong correlation (i.e. close to 1) and dark purple for a strong anti-correlation (i.e. close to $-1$). The panels along the x- and y-axis display the mean He~\textsc{i} and H$\alpha$ profiles, respectively, in black, with their variance in blue.}
    \label{fig:cmhahe}
\end{figure}

Furthermore, we computed a correlation matrix between H$\alpha$ and He~\textsc{i}.
This tool consists in the computation of a linear correlation coefficient between the velocity channels of two lines.
It allows to trace the physical processes dominating the spectral variation, without identifying it clearly.
A strong correlation coefficient between two regions of two spectral lines indicates that the variations of those regions are dominated by a common physical process.
In the opposite case, anti-correlation may highlight one process (with opposite implications at different velocities in the line) or two linked processes dominating the lines' variation.
The correlation matrix between the H$\alpha$ and He~\textsc{i} lines is presented in Fig. \ref{fig:cmhahe}.
This matrix exhibits a strong correlation between the NC of the He~\textsc{i} line and the blue wing of H$\alpha$ line, which was the part of the line showing the 5 d period.
We can also note a correlation of the broad component of the He~\textsc{i} line with the red wing of H$\alpha$. 
Those regions are both modulated on the orbital period, and should trace the accretion as well, they might be closely linked to the pulsed accretion phenomenon.

\subsection{Mass accretion rate}
\label{subsec:macc}

In order to derive the mass accretion rate of DQ Tau, we used the relationship between the accretion luminosity and the line luminosity \citep{Alcala17}
\begin{align}
    \log \left(L_{\rm acc}\right) &= a \log \left(L_{\rm line}\right) + b,
    \\
    L_{\rm line} &= 4 \pi d^{2} F_{\rm line},
    \\
    F_{\rm line} &= F_{\rm 0} \cdot EW \cdot 10^{-0.4 m_{\lambda}},
\end{align}

\noindent where $L_{\rm acc}$ and $L_{\rm line}$ are the accretion and line luminosity, respectively, $a$ and $b$ are the linear coefficients given by \cite{Alcala17}, $d$ is the distance from Gaia EDR3 \citep{Gaia21}, $F_{\rm line}$ is the line flux, $F_{\rm 0}$ is the reference flux at the wavelength corresponding to the line, $EW$ is the line equivalent width, and $m_{\lambda}$ is DQ Tau's magnitude in the selected wavelength corrected from the extinction.
The mass accretion rate is then given by the following
\begin{equation}
    L_{\rm acc} = \frac{G M_{\star} \dot{M}_{\rm acc}}{R_{\star}} \left[ 1 - \frac{R_{\star}}{R_{\rm t}}\right].
\end{equation}
Here $R_{\rm t}$ corresponds to the truncation radius. We took a typical value of $R_{\rm t}$~=~5~$R_{\star}$ \citep{Bouvier07b}, as it was used in previous works.
Rigorously, the value of $R_{\rm t}$ is dependent of the magnetic field strength and of the accreted material ram pressure. For binaries, this value is predicted to be phase-dependent \citep{Munoz16}, and higher at the periastron passage where the ram pressure is likely higher. Even if a fixed value is less physical, the mass accretion rate does not seem too sensitive to $R_{\rm t}$ variation, a decrease by a factor 2 yields a factor 0.8 in the mass accretion rate.

\begin{figure}
    \centering
    \includegraphics[width=.49\textwidth]{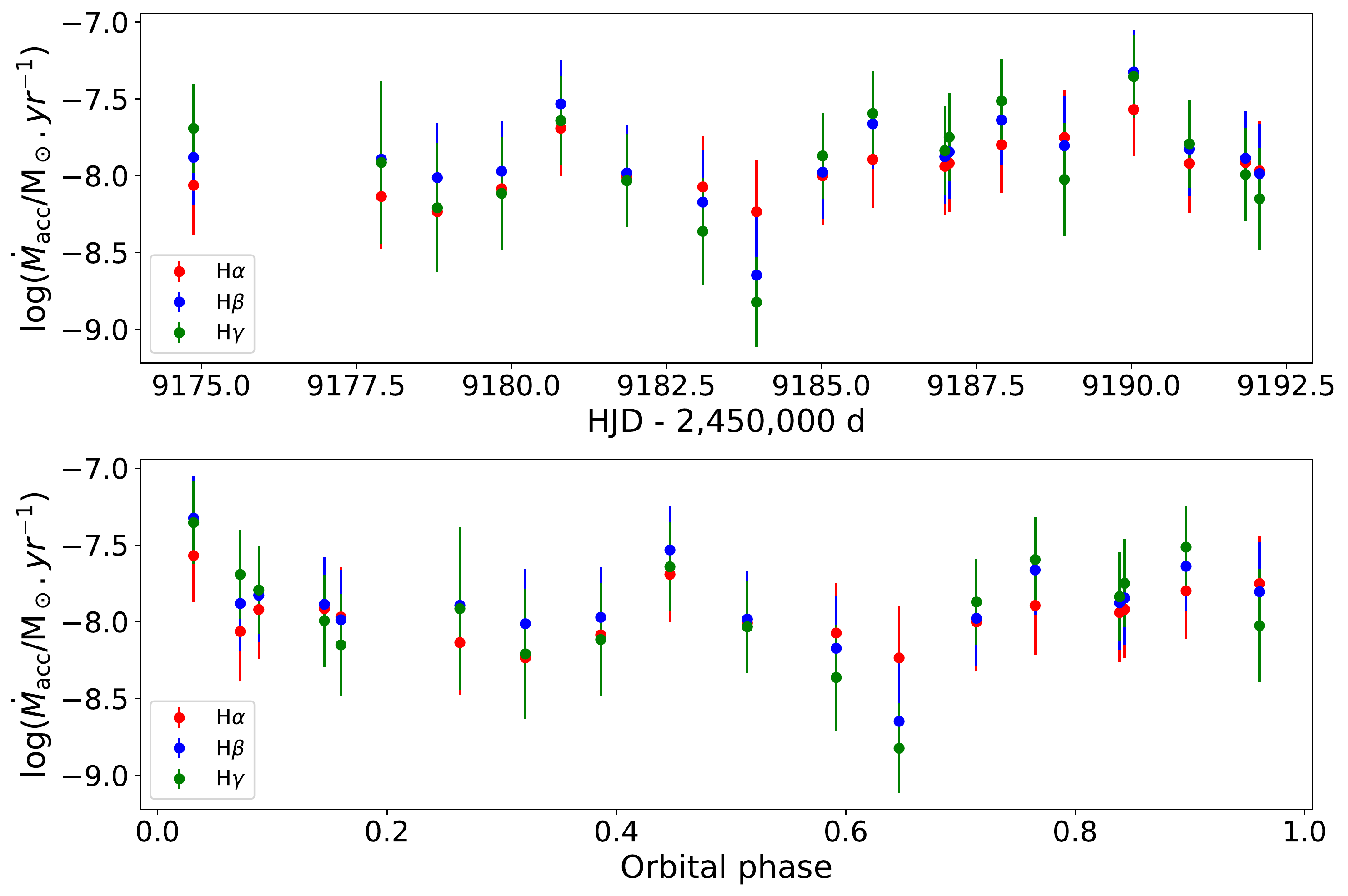}
    \caption{Mass accretion rate computed from H$\alpha$ \textit{(red)}, H$\beta$ \textit{(blue)}, and H$\gamma$ \textit{(green)} residual profiles as a function of HJD \textit{(top)} and the orbital phase \textit{(bottom)}.}
    \label{fig:maccBalmer}
\end{figure}

We derived the mass accretion rate from the three strongest lines of DQ Tau's spectrum, namely the residual H$\alpha$, H$\beta$, and H$\gamma$ lines. 
From these three lines, we obtained a mean mass accretion rate of about 10$^{-8}$ \msunyr, consistent with the values obtained by \cite{Fiorellino22} from XSHOOTER data. 
However, the variability of the mass accretion rate is different from their work, exhibiting two maxima at phases 0.03 and 0.45, corresponding to the periastron and apastron, respectively.
The corresponding curve is shown in Fig.~\ref{fig:maccBalmer}.
This is consistent with the photometric light curves studied by \cite{Tofflemire17} and \cite{Kospal18}, which indicates that accretion events occasionally happen near apastron as well.

\subsection{Zeeman-Doppler imaging}
\label{subsec:zdi}

In order to retrieve the large-scale magnetic field topology of the system's two components, we performed a complete Zeeman-Doppler imaging (ZDI) analysis using the LSD profiles shown in Figs. \ref{fig:stokesILSD} and \ref{fig:stokesVLSD}. 
The observation at HJD 2~459~177.9 exhibiting a low S/N, was excluded it from this analysis. 
We used the \texttt{inversLSDB} implementation of ZDI \citep{Rosen18}, which is a binary-star adaptation of the \texttt{inversLSD} code developed by \cite{Kochukhov14}.
This implementation was successfully used in previous studies of double-lined spectroscopic binaries \citep[see][for various examples]{Kochukhov19,Lavail20, Hahlin21}.

To compute the local Stokes I profile, we used the same procedure as in \citet{Pouilly21}.
We derived the LSD Stokes I profile of a slow rotator, with a spectral type close to DQ Tau components, using the same parameters as for DQ Tau's LSD profile derivation.
We choose HD~88230, a star with the spectral type K6 and a projected rotational velocity of \vsini\ = 1.8 \kms.
Finally, we found the Gaussian and Lorentzian widths by fitting a Voigt function to its LSD Stokes I profile. 
The resulting Voigt function parameters were used together with the mean wavelength and Land\'e factor determined earlier to describe the local Stokes I and V profiles in the framework of the Unno-Rachkovsky analytical solution \citep{polarization04} of the polarized radiative transfer.
We assumed both components of the DQ Tau system to have the same local Stokes I profile.
We used the orbital solution derived in this work (Table \ref{tab:orbElements}) and the stellar parameters from Table \ref{tab:param}, more precisely the \vsini, rotation period, and the inclination of both components (assuming both inclinations are the same).
Concerning the missing parameters of the secondary, we assumed the inclination equal to the primary's one, and we derived the secondary's rotation period by running the ZDI analysis on a grid on periods and using the minimal deviation of the Stokes V fit to indicate the optimal value, yielding \prot(B) = 3.01 d.

\begin{figure*}
    \centering
    \includegraphics[width=.49\textwidth]{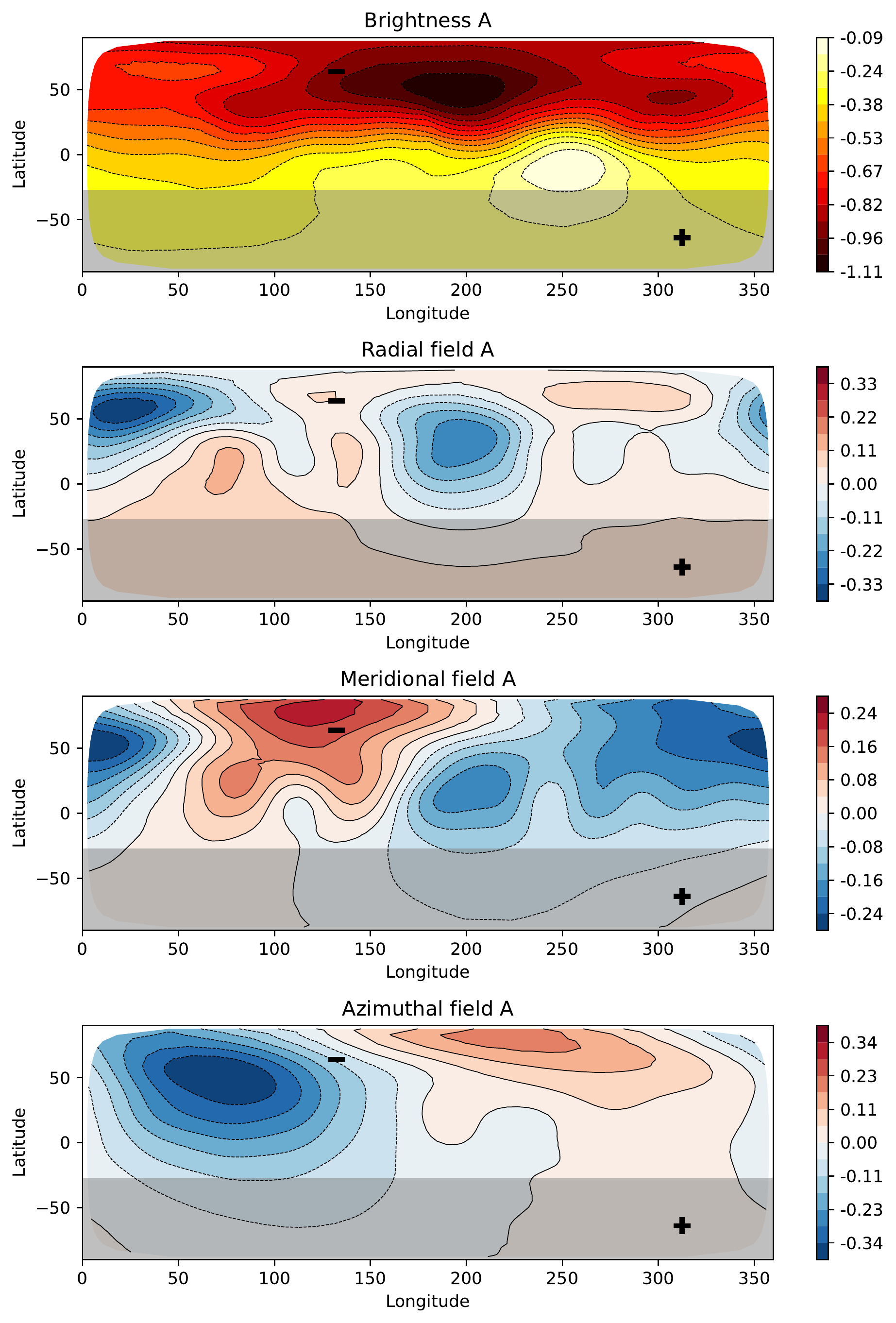}
    \includegraphics[width=.49\textwidth]{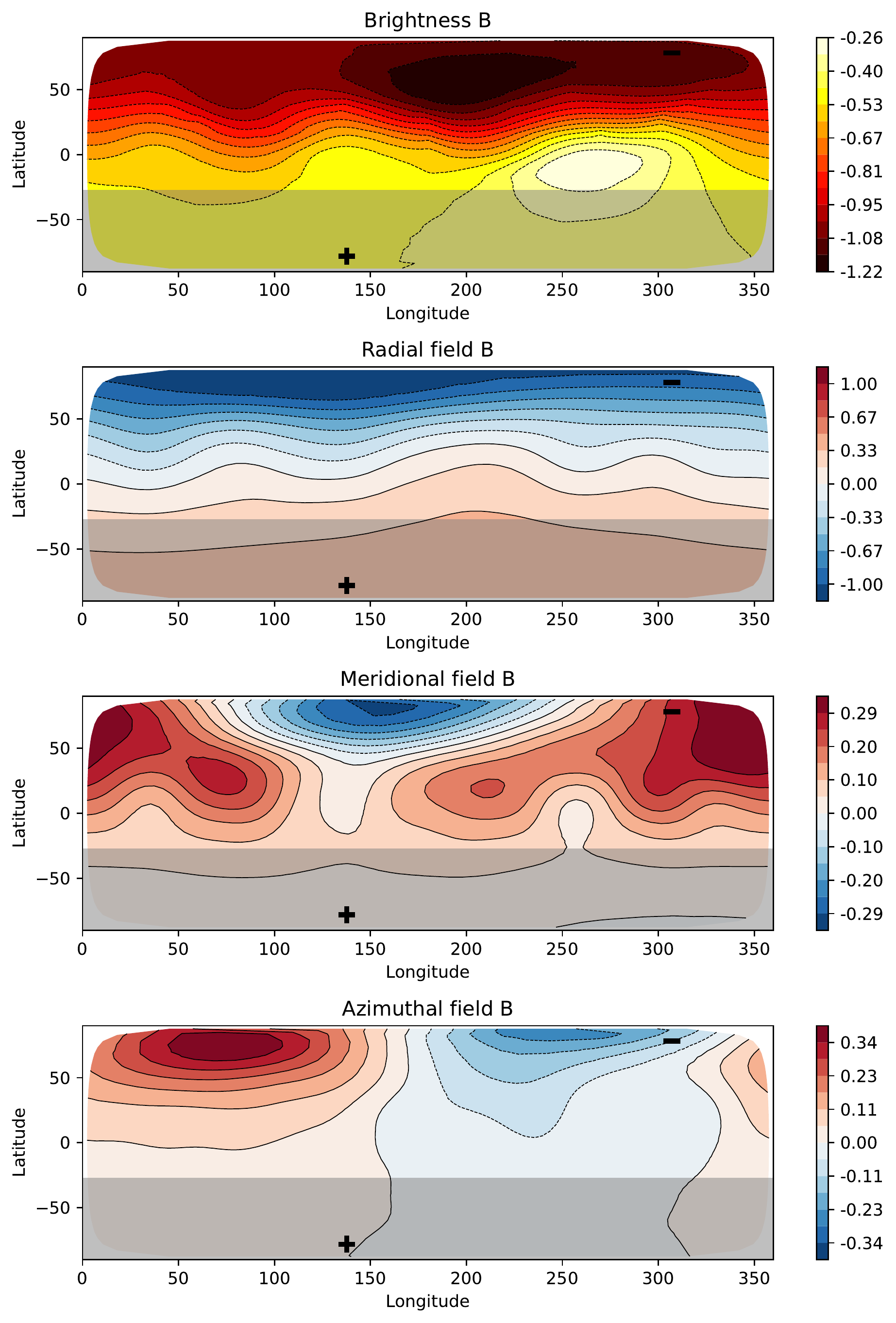}
    \caption{ZDI magnetic field maps of the A component \textit{(left)}, and B component \textit{(right)}.The 0-latitude represents the stellar equator, 90$^\circ$ and $-$90$^\circ$ the North and South rotational poles, respectively. The 0 to 360$^\circ$ longitudes scale the phases 0 to 1 of the stellar rotation cycle. The grey-shaded areas represent the unseen region of the stellar surface due to the stellar inclination given in Table \ref{tab:param}. The "+" and "-" signs are locating of the dipolar positive and negative pole, respectively. \textit{Top panels:} brightness maps, the color bar indicates the logarithm of the normalised brightness. \textit{Three bottom panels:} radial, meridional and azimuthal fields. The color bar is in units of kG.}
    \label{fig:zdimaps}
\end{figure*}

The brightness map is reconstructed by modeling the Stokes I profiles, and is shown in Fig. \ref{fig:zdimaps}.
Both components are showing similar brightness structure, with a major dark spot around 50$^\circ$ in latitude and 200$^\circ$ in longitude.
This map is then used as an input to reconstruct the magnetic maps, modelling the Stokes V profiles.
The magnetic field is defined as a combination of a poloidal and a toroidal field, decomposed using spherical harmonics.
We set the maximum degree of the spherical harmonic coefficients to $\ell_{\rm max}$=10, which restricts the complexity of the magnetic topology as the moderate \vsini\ of both components of the system allows to retrieve only the larger structures, and thus the lower order of the magnetic field topology.
The resulting brightness and magnetic maps are shown in Fig. \ref{fig:zdimaps} and the fit to the LSD profiles is illustrated in Fig. \ref{fig:ZDIfit}.

The extended list of magnetic characteristics resulting from this analysis is provided in Table \ref{tab:magTop}.
Both components' magnetic energy is dominated by the poloidal field, itself largely dominated by the dipolar component ($\ell$=1), with a significant quadrupolar contribution ($\ell$=2).
The primary is exhibiting a mean surface field of $\langle B \rangle$ = 0.163 kG with a maximum field modulus reaching 0.551 kG. The secondary is showing larger values, with $\langle B \rangle$ = 0.568 kG and $B_{\rm max}$ = 1.322 kG.
The dipolar negative pole location, derived from the maximum of the radial component of the dipolar magnetic field only, seems to be located at about the latitude 64$^\circ$ and longitude 132$^\circ$ on the A component, reaching 230 G and at about latitude 78$^\circ$ and longitude 307$^\circ$ on the B component, reaching 621 G.

\begin{table}
    \centering
    \caption{Large-scale magnetic characteristics of the DQ Tau system resulting from the LSD Stokes V profile analysis.}
    \begin{tabular}{l l l}
    \hline
        Distribution of & A  & B \\
        the magnetic energy & (\%) & (\%) \\
        \hline
        $\ell$=1 & 67.8 & 82.0 \\
        $\ell$=2 & 17.9 & 9.9 \\
        $\ell$=3 & 6.1 & 3.4 \\
        $\ell$=4 & 4.4 & 1.1 \\
        $\ell$=5 & 2.1 & 0.4 \\
        $\ell$=6 & 0.8 & 0.1 \\
        $\ell$=7 & 0.5 & 0.0 \\
        $\ell$=8 & 0.3 & 0.0 \\
        $\ell$=9 & 0.1 & 0.0 \\
        $\ell$=10 & 0.1 & 0.0 \\
        poloidal field & 67.1 & 96.1 \\
        toroidal field & 32.9 & 3.9 \\
        axisymmetry (|$m$|<$\ell$/2) & 32.3 & 89.2 \\
        \hline
        Magnetic field strengths & (kG) & (kG) \\
        \hline
        $\langle B \rangle$ & 0.163 & 0.568 \\
        $B_{\rm max}$ & 0.551 & 1.322 \\
        $B_{\rm dip,max}$ & 0.230 & 0.621 \\

    \hline
    \end{tabular}
    \label{tab:magTop}
\end{table}

\subsection{Zeeman intensification}
\label{subsec:zeemanInt}

The small-scale magnetic field, which contains the major part of cool stars' magnetic energy and is neglected by ZDI, induces a change in the equivalent width of magnetically sensitive lines \citep[e.g.][]{Kochukhov20}.
This effect is called Zeeman intensification and has a significant advantage for magnetic diagnostic of cool stars because it does not rely on the effect of magnetic field on line shapes (i.e. Zeeman broadening) for extracting information about the field.
This method, thus, places no requirements of a slow stellar rotation, very high spectral resolution and high S/N, typical of Zeeman broadening applications.  
It is thus possible to use Zeeman intensification for magnetic studies using moderate-quality observations at optical wavelengths, as done for the ZDI analysis. This allows one to determine characteristics of both small- and large-scale fields from the same data set.

To perform the Zeeman intensification analysis, we used the algorithm developed by \citet{Hahlin21}.
The code written in IDL use the \texttt{SoBAT} library \citep{Anfinogentov21} to perform a Markov Chain Monte Carlo (MCMC) sampling on a grid of synthetic spectra produced by the \texttt{SYNMAST} code, a polarized radiative transfer code described by \citet{Kochukhov10}.
These calculations are based on a line list retrieved from the VALD database \citep{Ryabchikova15}. Furthermore, we assumed solar chemical abundances \citep{Asplund09} and used MARCS model atmospheres \citep{Gustafsson08}. 
Finally, we used a uniform radial magnetic field, parametrised as a sum of magnetic field strength ranging from 0 to 6 kG, with a 2 kG step, weighted by filling factors representing the amount of stellar surface covered by this magnetic field.

As described in \citet{Hahlin22}, the 963.5 to 981.2 nm region is highly suitable for Zeeman intensification analysis since it contains a group of 10 Ti~{\sc i} lines with different magnetic sensitivity, including the magnetic null line at 974.3 nm.
This allowed us to disentangle the effect of magnetic field on the equivalent widths from the effect of any other parameters, such as the Ti abundance, by fitting simultaneously magnetically sensitive and insensitive lines.
The results of the disentangling procedure are presented in Fig.~\ref{fig:disTi}.

The first step was thus to disentangle the spectra of binary components, for which we used the procedure described in Sec.~\ref{subsec:specdis}, slightly modified because of the wavelength region used.
Indeed, the 963.5 to 981.2 nm region is contaminated by variable telluric absorption, which needs to be modelled and removed. To this end, we applied a modified version of the disentangling routine presented in Sec.~\ref{subsec:specdis}, described by \citet{Kochukhov19}.
This version allowed to separate the three components of the spectra, meaning A, B, and telluric components.
The observations are thus assumed to be the superposition of three spectra: primary's and secondary's spectra, as previously, and the telluric spectrum, shifted by the heliocentric velocity correction and scaled for each observation by a power-law.

Then, for each binary component, we produced a grid of \texttt{SYNMAST} synthetic spectra, spanning a suitable range of magnetic field strengths and Ti abundances.
We adopted \teff\ and \vmic\ from Table \ref{tab:param}, and \logg\ = 4.0.

\begin{figure*}
    \centering
    \includegraphics[width=.45\textwidth]{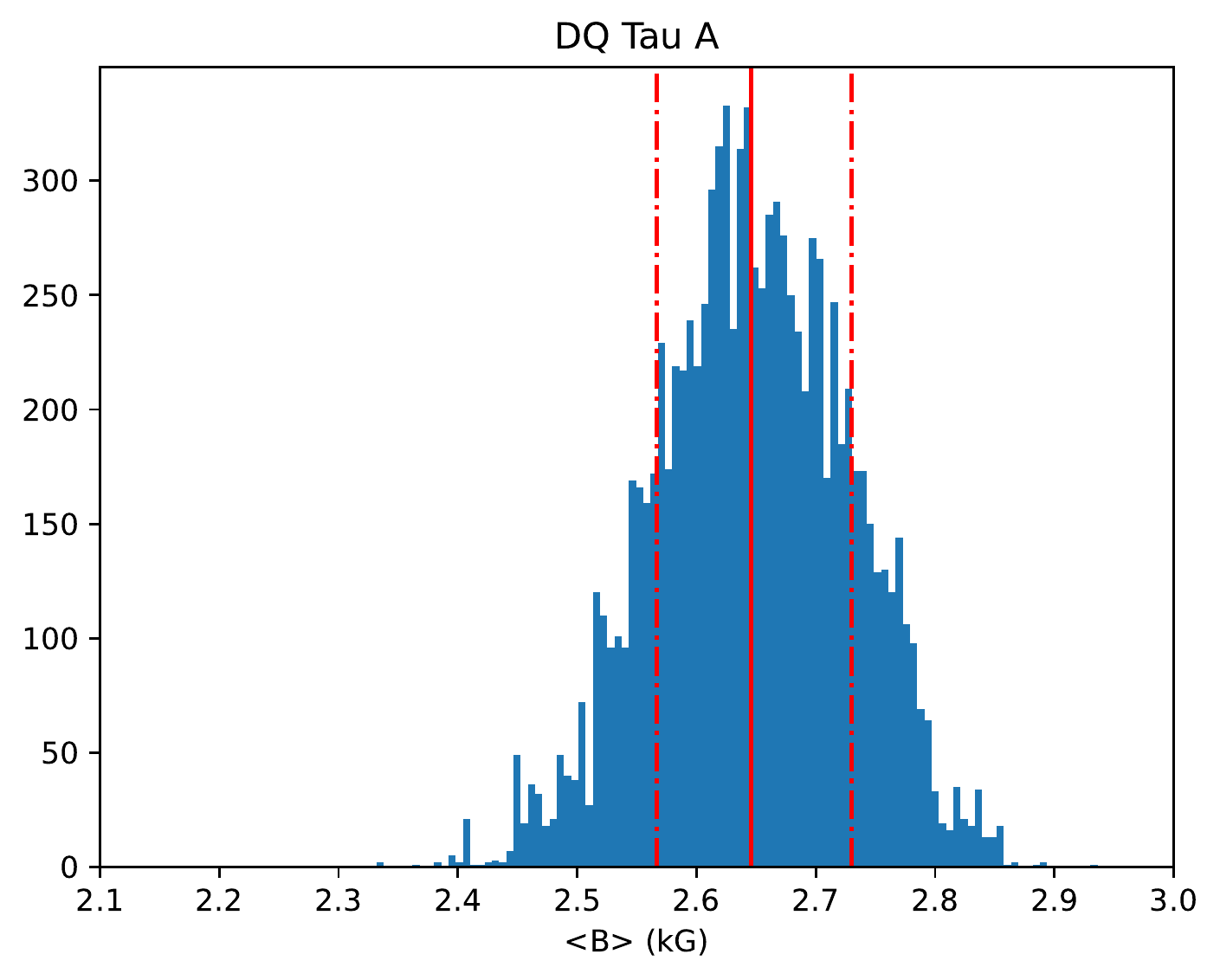}
    \includegraphics[width=.45\textwidth]{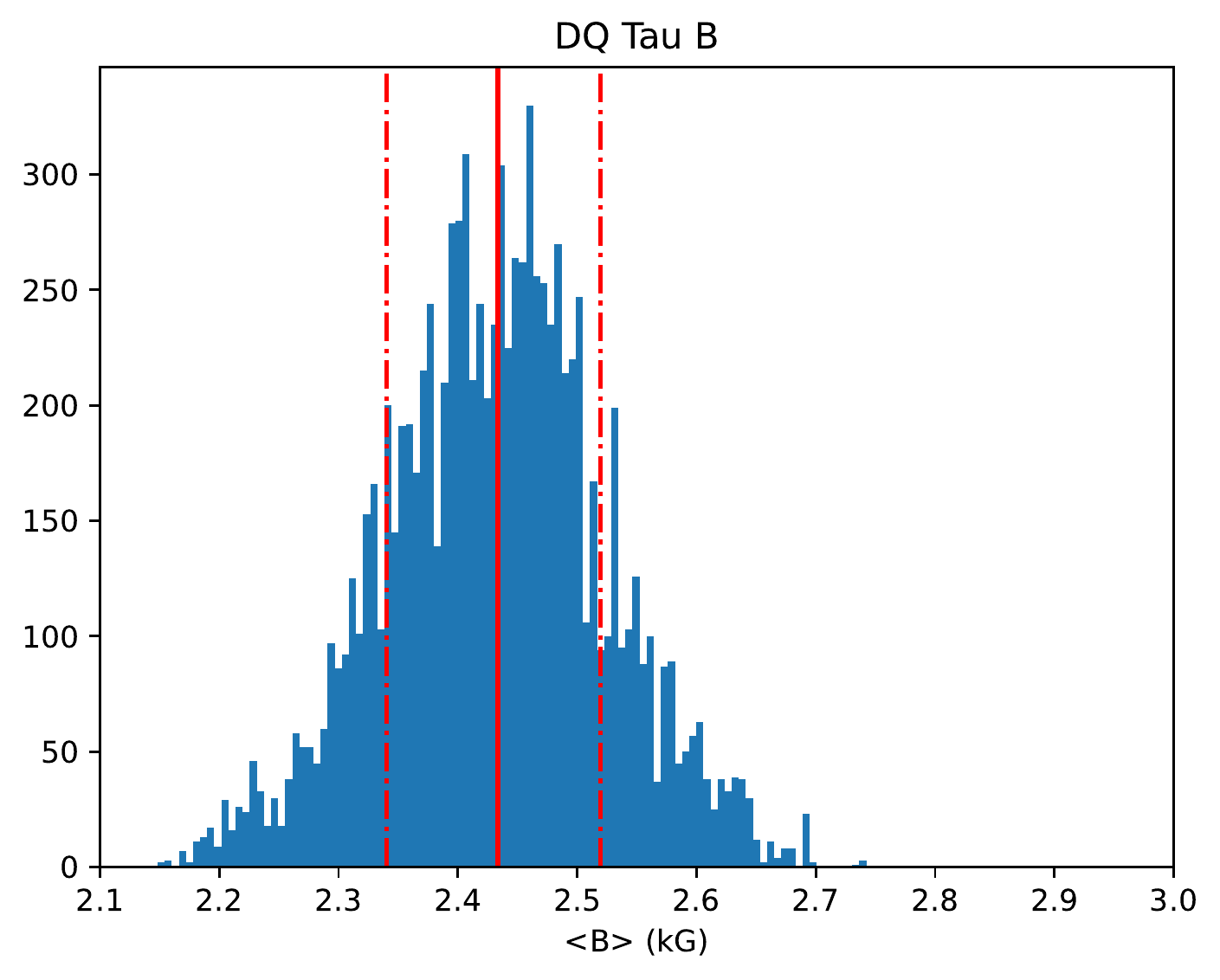}
    \caption{Posterior distribution of the surface-averaged magnetic field strength from the Zeeman intensification analysis for A \textit{(left)} and B \textit{(right)} components of DQ Tau. The solid vertical line represents the median value and the dash dotted ones represent the 68 per cent confidence intervals.}
    \label{fig:Bfield}
\end{figure*}

\begin{figure*}
    \centering
    \includegraphics[width=.49\textwidth]{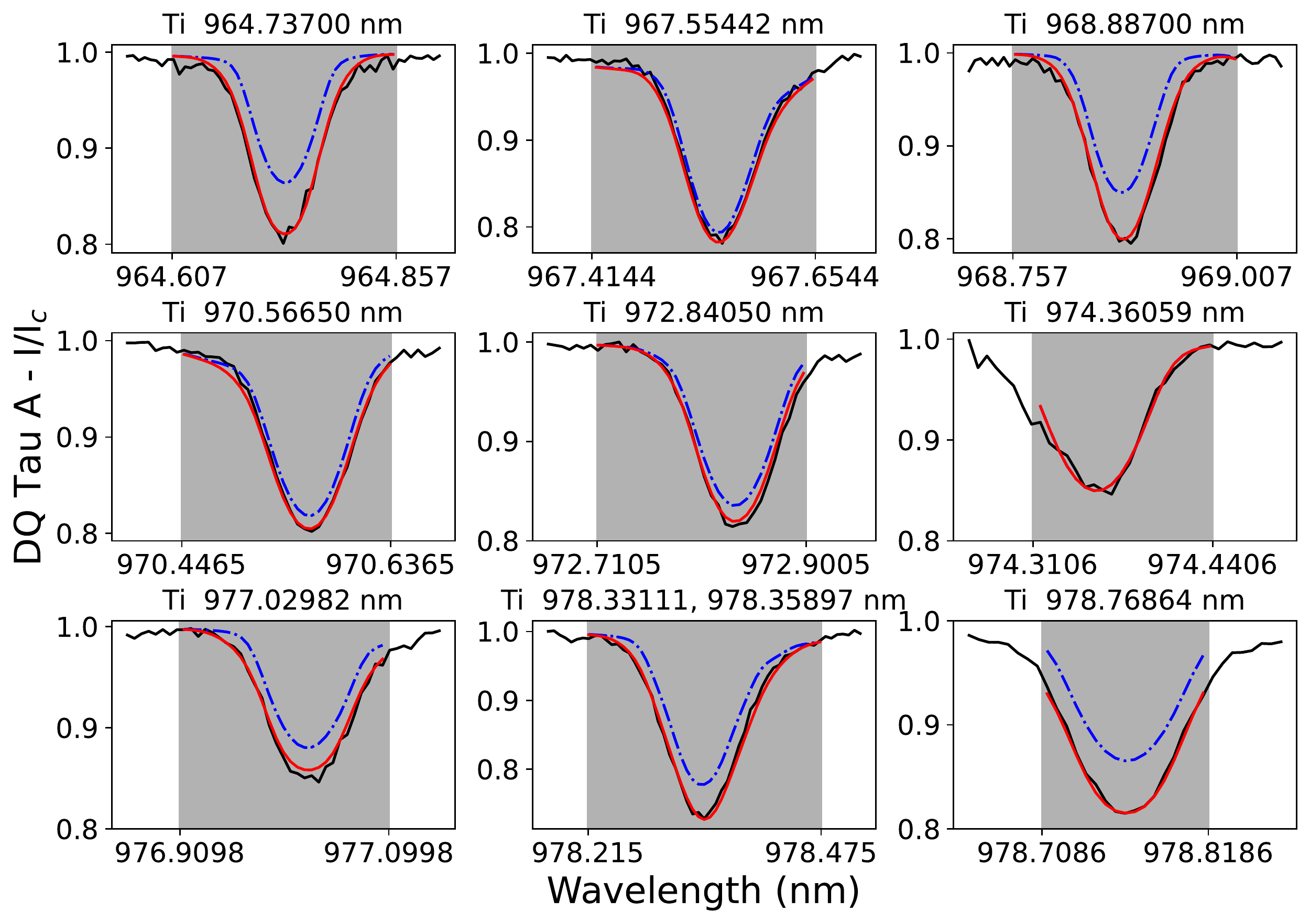}
    \includegraphics[width=.49\textwidth]{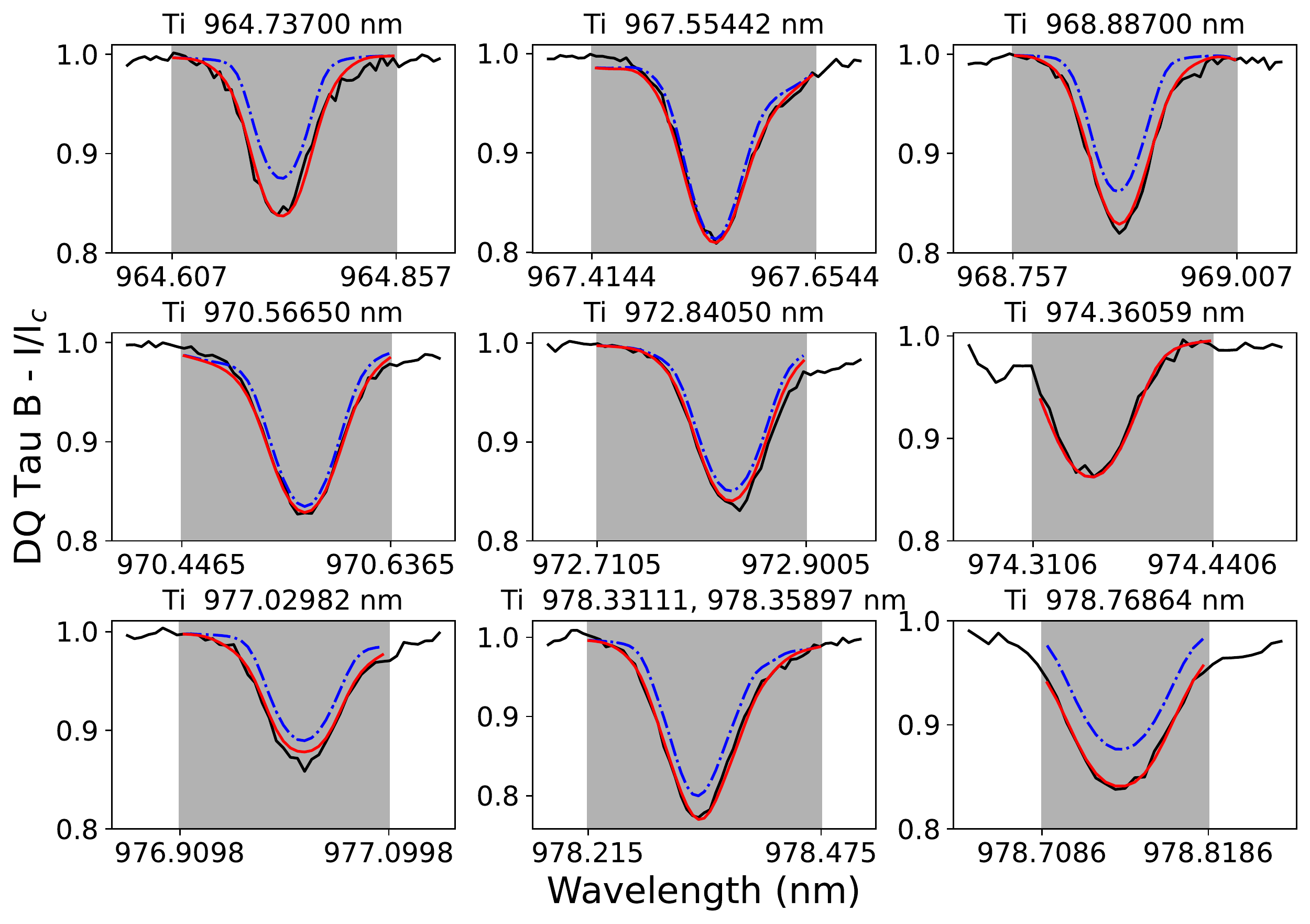}
    \caption{Observed Ti~{\sc i} lines used for the Zeeman intensification analysis \textit{(black)} and the spectrum synthesis fits corresponding to the optimal parameters \textit{(red)} for A \textit{(left)} and B \textit{(right)} components. The dashed blue lines correspond to the spectra without magnetic field. The grey area represents the region where the fit is performed and the central wavelength of each line is given at the top of each panel.}
    \label{fig:Tilines}
\end{figure*}

Figure \ref{fig:Bfield} shows the posterior distribution of the surface-averaged magnetic field strength. The fit to the mean Stokes I profiles of the Ti~{\sc i} lines is presented in Fig. \ref{fig:Tilines}.
According to this analysis, DQ Tau A and B exhibit similar mean magnetic field strengths, $2.65 \pm 0.08$ kG and $2.43 \pm 0.09$ kG, respectively.
The full posterior distribution of the parameters derived by this procedure is shown in Fig.~\ref{fig:corner}.

\section{Discussion}
\label{sec:discuss}

DQ Tau was selected for a study of the magnetospheric accretion process along an orbital cycle because of its very suitable properties for a spectropolarimetric monitoring.
The orbital period of this system is around 16 d, and the rotation period of the primary is about 3 d, allowing one to obtain observations during a complete orbit and keeping a good sampling of the stellar rotation cycle.
Furthermore, DQ Tau is composed of two CTTSs, for which the accretion was studied for many single objects, but rarely for binary systems.
Finally, the eccentricity of the system (about 0.6), yields a separation at periastron small enough to allow an interaction between the magnetospheres of both components.
This is revealed by strong bursting events in the Kepler K2 light curve at each periastron passage, visible on the more recent TESS light curve as well.
The aim of this work is thus to investigate the processes responsible for those events, likely linked to the accretion as mass accretion rate enhancement has been reported over past cycles \citep{Tofflemire17, Kospal18, Fiorellino22}.
We will thus focus on the magnetospheric accretion process for both components, and study how it is disturbed around the periastron and along the orbital cycle.\\

\subsection{At orbital scale}
\label{subsec:orbScale}

\begin{figure}
    \centering
    \includegraphics[width=.49\textwidth]{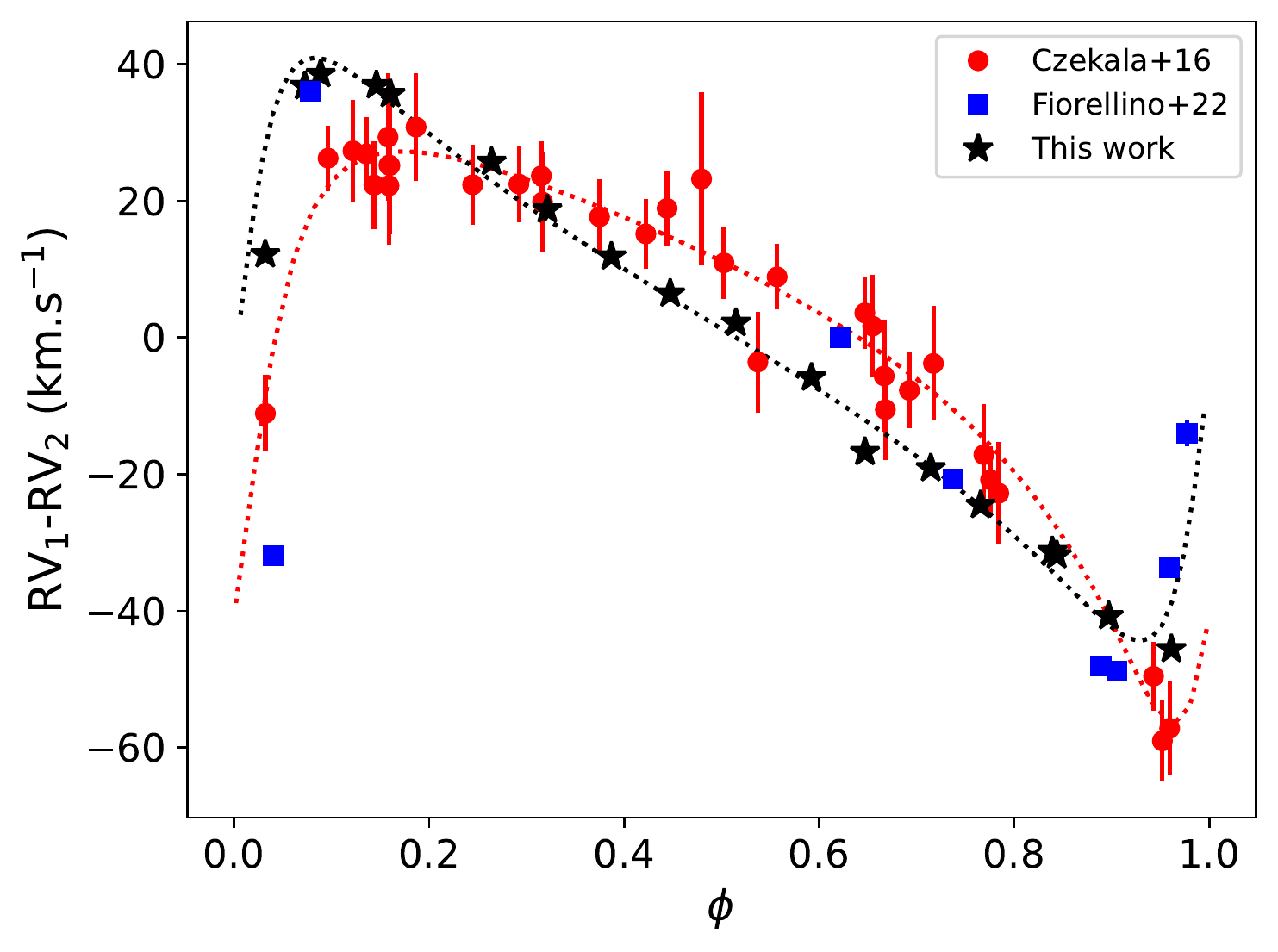}
    \caption{Velocity difference between the primary and the secondary as a function of the orbital phase. The symbols indicate the corresponding publication and the error bars may be smaller than the symbol's size. The dotted black line is the orbital solution derived in this work. The one in red is the orbital solution found by \citealt{Czekala16}.}
    \label{fig:vradCzekala16}
\end{figure}

The spectropolarimetric time-series used in this work consists of 18 observations covering a complete orbital cycle.
From the LSD Stokes I profiles, we derived the radial velocities of both components for each observation.
Most of the orbital elements derived from these radial velocity curves are consistent with the values found by \cite{Czekala16} within the uncertainties, excepting the argument at periastron $\omega$, changing by $\sim$35$^\circ$ (see Table \ref{tab:orbElements}).
Furthermore, the slope of the velocity difference of the two components between two periastron passages seems steeper than the one found by \cite{Czekala16} (see Fig. \ref{fig:vradCzekala16}), which is consistent with a change of $\omega$.
This indicates apsidal motion, meaning that the orbital ellipse rotated within the orbital plan of the system between the observations used by \cite{Czekala16}, taken between 1984 and 1994, and the 2020 ESPaDOnS time series used in this work.
The results obtained by \cite{Fiorellino22} from the observations in 2012--2013 seem to confirm this trend, as they determined velocities at 8 epochs between the measurements by \cite{Czekala16} and our work.
A fit of the argument at periastron of the data set used by \cite{Fiorellino22} (assuming other orbital elements derived in this work) yields $\omega$ = 256 $\pm$ 7$^\circ$.
This is an evidence of the apsidal motion with a rate of $\dot{\omega}$ = 1.15 $\pm$ 0.25$^\circ$ yr$^{-1}$, depending on whether we use 1984 or 1994 as the starting date.
This phenomenon usually occurs in the tidally locked close binary systems, where tidal interactions between the two components will distort their shapes and thus their gravitational potential. 
In the case of an eccentric orbit, this distortion will induce a precession of the orbit -- the apsidal motion \citep[][and references therein]{Feiden13}. 
DQ Tau is not a particularly close binary system and shows no evidence of being tidally locked.
Nevertheless, its high eccentricity and small separation at periastron may be responsible for such gravitationally-induced phenomena. 
Even if we cannot link the apsidal motion to the magnetospheric interactions, this is the first time that we see signs of apsidal motion in DQ Tau, and that in such a young, accreting system with strong magnetic fields. There may be multiple causes for it, which may be the gravitational interaction of the stars at periastron, the accretion of material from the circumbinary disk, and tidal interaction as well.

\begin{figure}
    \centering
    \includegraphics[width=0.4\textwidth]{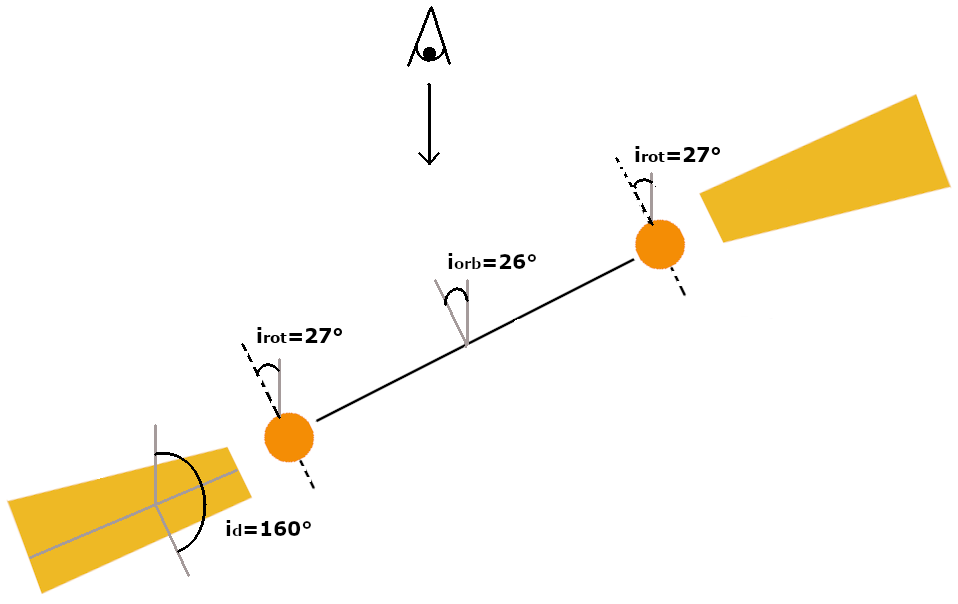}
    \caption{Sketch representing the inclination angles of the components' rotation axis, the orbit rotational axis, and the disc's plane as used in this work. The eye with the arrow indicate the line of sight, the orange circles represent the two components and the dashed lines their rotation axis. The black line between them is the orbital plan and the disc is represented by the yellow part on each side. The grey lines are guiding the angle visualisation.}
    \label{fig:sketchInc}
\end{figure}

The mass accretion rate computed from the residual H$\alpha$, H$\beta$, and H$\gamma$ lines exhibits the expected modulation on the orbital period, with the two maxima, at periastron and apastron.
This is consistent with the Kepler K2 light curve, showing strong bursting events at each periastron passage and occasionally at the apastron as well.
However, even if the NC amplitude of the He~\textsc{i} line at 587.6 nm is maximal near the periastron passage, the one around apastron is quite weak.
This may mean that the accretion shock is not visible at the moment of the observation, or another process is governing the emission of Balmer lines at this moment, independently of the main accretion funnel flow responsible for the accretion shock and the NC of the He~\textsc{i} line.
Nevertheless, the inclination of the stars' rotational axis, $i$ = 27$^{\circ}$ and of the orbit's rotational axis, $i_{\rm orb}$ = 26$^{\circ}$ (derived from the orbital solution and the stellar atmospheric parameters), indicate an alignment with the circumbinary disk \citep[$i_{\rm d}$ $\sim$ 160$^{\circ}$ from][]{Czekala16}.
This may allow an interaction between the stars and the disk at apastron, which can be at the origin of the apastron's burst.
A sketch displaying the different inclination angles is provided in Fig. \ref{fig:sketchInc}.

Following the method of \cite{Fiorellino22}, we identified the main accretor at each epoch using the velocity of the highest emission peak in the line.
From the NC of He~\textsc{i} line at 587.6 nm, the B component is the main accretor during most of the orbital cycle, between phases 0.59 and 0.96, meaning from right after the apastron passage until the periastron passage.
Between phase 0.03 and 0.51 it is less well defined.
Depending of the epoch, it is shifting between the two components, or the profile does not allow to clearly identify the main accretor.
The H$\alpha$ line is showing another, clearer behaviour. Analysis of this line suggests the B component as the main accretor from phase 0.9 to phase 0.39, and then switching for the A component as the main accretor from phase 0.51 to phase 0.84.
Following the interpretation by \cite{Fiorellino22}, the H$\alpha$ and the NC of the He~\textsc{i} line at 587.6 nm are tracing different places in the accretion stream.
The NC is formed in the post-shock region of the accretion spot \citep{Beristain01} whereas the H$\alpha$ is formed in the accretion funnel flow \citep{Muzerolle01}.
The difference in the main accretor between the two lines can thus be interpreted as if the H$\alpha$ information gives a prediction of the He~\textsc{i}, as the accretion starts in the funnel flow before reaching the stellar surface.
The Table \ref{tab:mainAcc} summarises this analysis for H$\alpha$, H$\beta$, H$\gamma$, Ca~\textsc{ii} IRT, and the He~\textsc{i} lines.\\

\begin{table}
    \centering
        \caption{Velocity position of the highest emission peak in H$\alpha$, H$\beta$, H$\gamma$, Ca~\textsc{ii}, and NC of He~\textsc{i} lines used as identifier of the main accretor at each orbital phase. "A" and "B" indicate the primary or the secondary as the main accretor according to the given line, "?" indicates that we were not able to identify the main accretor from the given line at that phase.}
    \begin{tabular}{|l|c|c|c|c|c|}
    \hline
    $\phi_{\rm{orb}}$ & H$\alpha$ & H$\beta$ & H$\gamma$ & Ca~\textsc{ii} &  He~\textsc{i} \\
    \hline
    0.03 & B & B & B & B & A \\
    0.07 & B & B & B & B & B \\
    0.09 & B & A & B & B & A \\
    0.15 & B & A & A & A & B \\
    0.16 & B & B & B & A & B \\
    0.26 & B & ? & ? & ? & ? \\
    0.32 & B & ? & B & A & A \\
    0.39 & B & ? & B & A & ? \\
    0.45 & ? & ? & ? & ? & ? \\
    0.51 & A & ? & ? & ? & ? \\
    0.59 & A & A & A & ? & B \\
    0.65 & A & B & ? & B & B \\
    0.71 & A & A & A & B & B \\
    0.76 & A & A & A & B & B \\
    0.84 & A & B & B & B & B \\
    0.84 & A & B & B & B & B \\
    0.90 & B & B & B & B & B \\
    0.96 & B & B & B & A & B \\
        
    \hline
    \end{tabular}
    \label{tab:mainAcc}
\end{table}

\subsection{At stellar scale}
\label{subsec:stellarScale}

Here, we also studied the two components independently, focusing on their atmospheric and fundamental parameters, and magnetic field characteristics.
We disentangled the spectra of the primary and the secondary using the method described in Sec. \ref{subsec:specdis}.
This allowed us to derive the atmospheric parameters for each component and study the Zeeman intensification of their near-infrared Ti~{\sc i} lines (see Sec. \ref{subsec:zeemanInt}). 
The latter analysis yielded information on the small-scale magnetic fields of both stars.
With the help of the ZDI method (Sec. \ref{subsec:zdi}), we reconstructed the brightness and magnetic field maps from the composite LSD profiles, allowing us to analyse the magnetic field topologies of DQ Tau components.

The stellar parameters obtained from the fit of the \texttt{ZEEMAN} synthetic spectra and then the HRD position fit are typical of low-mass CTTSs with a K7--K8 spectral type and a mass of $\sim$0.6 M$_\odot$ (Table \ref{tab:param}) for both components.
However, the B component appears to rotate slower, indicating a slightly longer rotation period (up to 4.9 days) than for the primary if we assume a common inclination angle $i$ = 27$^\circ$.
This is consistent with the periodicity seen in the variation of the H$\alpha$, H$\beta$ and the NC of the He~\textsc{i} 587.6 nm line for which the 2D periodograms show a signal at a period slightly lower than 5 days.
Assuming that the NC of the He~\textsc{i} line is mainly emitted at the secondary’s velocity and traces the accretion shock at the stellar surface, this signal can be related to the rotation period of the secondary.
Even if the spectroscopic diagnostics point to a longer period for the secondary, the ZDI analysis points to another value.
Indeed, the deviation of the Stokes V fit seems lower for a 3-day period, closer to the primary's rotation period.
Nevertheless, a $\sim$5-day period might not be recovered by the ZDI analysis, as it would mean that the ESPaDOnS observations are covering only 3 rotation cycles, which might be too few to be disentangled from the orbital modulation of the profiles.

As discussed in Sec. \ref{subsec:orbScale}, we noticed the presence of emission peaks at the primary’s velocity at some phases in H$\alpha$, H$\beta$, H$\gamma$, Ca~\textsc{ii} and He~\textsc{i}.
This confirm that this component is also accreting, even if the main accretor seems to be mostly the B component.

The ZDI analysis reveals a large, dark, polar feature. 
Such large dark spots at the rotational poles are commonly seen on CTTSs and are consistent with the predictions of \cite{Kospal18} based on the quasi-sinusoidal light curve shape of DQ Tau.
The large-scale magnetic topology is mainly dipole, as expected for this kind of objects \citep{Kochukhov21}. However, in this case the dominance of the dipole component is massive.
The B component exhibits a stronger magnetic field than the A component (1.3 kG and 0.6 kG, respectively).
However, previous studies showed that ZDI does not completely recover the magnetic field strengths \citep[e.g.,][]{Lavail19, Kochukhov20}. 
To this end, we used a complementary Zeeman intensification analysis to estimate the total mean magnetic field strengths for both components.
Even if the B component has a large-scale field greater than the A component, both small-scale fields are similar (about 2.5 kG). 
This means that about 23 per cent of the secondary's magnetic flux is recovered by ZDI, but only 6 per cent of the primary.
This difference in the ZDI ability to recover the magnetic field strengths may reside in the axisymmetry of the fields of both components.
The contribution of the axisymmetric component is weaker in the primary's magnetic topology (32.3 per cent against 89.2 per cent for the secondary). 
This is in agreement with previous studies \citep{Lavail19, Lavail20, Hahlin22}, showing that ZDI recovers a larger fraction of the magnetic field flux for simple and axisymmetric field geometries.
From the physical point of view, the surface field strengths we derived are perfectly consistent with the two optical flares detected by \cite{Tofflemire17}, which could happen if the surface magnetic field is reaching 1.5 kG.
Furthermore, these field strengths, in addition to an interaction between the two magnetospheres at periastron, may accelerate electrons to relativistic speeds resulting in synchrotron radiation
and the millimetre flares as well \citep{Salter10}.

\section{Conclusions}
\label{sec:ccl}

DQ Tau is one of the few known spectroscopic binary systems composed of accreting components.
The accretion processes play a key role in stellar evolution. As most of the low-mass Sun-like star are born in multiple systems, studying such systems is essential to improve the knowledge of stellar evolution.
Furthermore, DQ Tau is a short period ($P_{\rm orb}$ = 15.8 d), eccentric ($e \sim$ 0.6) binary composed of two accreting CTTSs (for which the accretion is magnetically driven), making it a perfect target for a spectropolarimetric monitoring.

Our data set is covering slightly more than one orbital cycle with a sampling of about one spectrum per night, which is ideal to study the accretion and the magnetic field along the stellar rotational and orbital cycle of both components.
The radial velocities measured along this orbital cycle showed an apsidal motion, a precession of the orbital ellipse, revealed by an increase of the argument of periastron between the value reported in the literature for past orbital cycles and the one obtained in this work. 
This phenomenon is common in tight or tidally locked binary systems, which is not the case for DQ Tau. 
We hypothesise that the explanation of apsidal motion is related to the eccentricity of the system, results in a short separation at periastron, making the tidal forces strong enough, and thus the gravitational potential distorted enough, to induce the observed apsidal motion in DQ Tau

We confirm in this work that both components of DQ Tau are accreting material, with a main accretor changing along the orbital motion and with a mean mass accretion rate of about 10$^{-8}$ \msunyr.
Furthermore, the mass accretion rate is modulated over the orbital period, with an enhancement of about an order of magnitude at periastron and apastron, suggesting interactions between the two components as well as between the stars and the circumbinary disc.
We derived the magnetic field structure of both components for the first time using ZDI to probe the large-scale fields and Zeeman intensification to probe the small-scale fields.
The magnetic topology, largely dominated by the dipole component, is compatible with a disc truncation and an accretion through funnel flows, as the higher-order components of the magnetic field are decaying quicker with the distance, the dipole one only is remaining at the disc scale.
The field strengths we derived are able to explain the phenomena previously observed for this system such as the optical and millimetre flares.

Despite we improved significantly our understanding of the accretion processes of this system, including the derivation of the first ever magnetic maps and magnetic field strengths, DQ Tau remains an enigmatic system which needs further investigations.
The origin of the apsidal motion as well as the strong mass accretion rate enhancements at periastron remain to be explained. 
The gravitational interaction model is a scenario that could describe both effects, but is limited because does not take into account the magnetic field which drives the accretion of such objects.
Finally, the study of one orbital cycle has improved our understanding of the system and provided us with a first glance of the effects due to the interaction of magnetospheres. 
However, to fully understand the dynamics of the interaction, monitoring on a larger time scale is necessary.
We thus plan to observe DQ Tau second time using the same instrument.
These observations should be carried out soon at CFHT.
This new time-series would provide an additional orbital cycle to compare with the existing observations and will improve the sampling around the periastron and apastron, helping to determine what is happening at these moments. In addition, these data will help to probe long-term evolution of the magnetic fields on both stars.

\section*{Acknowledgements}
We thank the anonymous referee for the comments provided, which drastically improved the clarity of this article.

Based on observations obtained at the Canada–France– Hawaii Telescope (CFHT) which is operated from the summit of Maunakea by the National Research Council of Canada, the institut National des Sciences de l’Univers of the Centre National de la Recherche Scientifique of France, and the University of Hawaii. The observations at the Canada–France–Hawaii Telescope were performed with care and respect from the summit of Maunakea which is a significant cultural and historic site.

O.K. acknowledges support by the Swedish Research Council (grant agreement no. 2019-03548), the Swedish National Space Agency, and the Royal Swedish Academy of Sciences.

This project has received funding from the European Research Council (ERC) under the European Union's Horizon 2020 research and innovation programme under grant agreement No 716155 (SACCRED).

\section*{Data Availability}

The spectropolarimetrc data used in this work are available from the Canadian Astronomy Data Center (CADC, \url{https://www.cadc-ccda.hia-iha.nrc-cnrc.gc.ca/en/}), program ID 20BF12.



\bibliographystyle{mnras}
\bibliography{mnras} 




\appendix

\section{\texttt{ZEEMAN} synthetic spectra fitting}

This section presents the results of the \texttt{ZEEMAN} synthetic spectra fitting routine to estimate the atmospheric stellar parameters of DQ Tau's 2 components on one the wavelength window. The analysis was done on 9 wavelength regions: 489.5-495.5, 522.0-526.0, 547.0-557.0, 592.0-598.0, 610.0-620.0, 639.5-649.0, 651.0-655.0, 658.0-668.0, and 744.0-754.0 nm.

\begin{figure*}
    \centering
    \includegraphics[width=.80\textwidth]{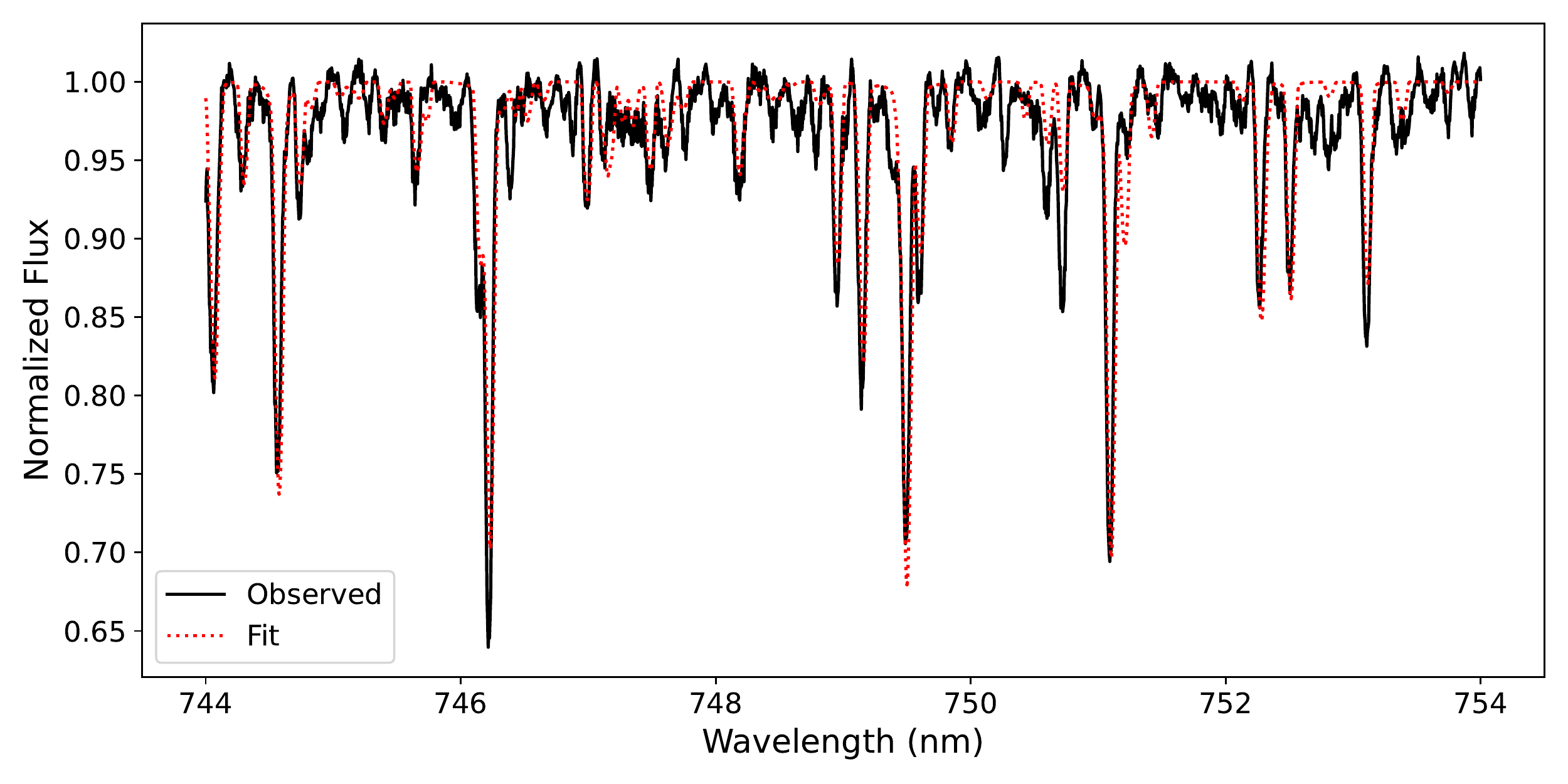}
    \includegraphics[width=.80\textwidth]{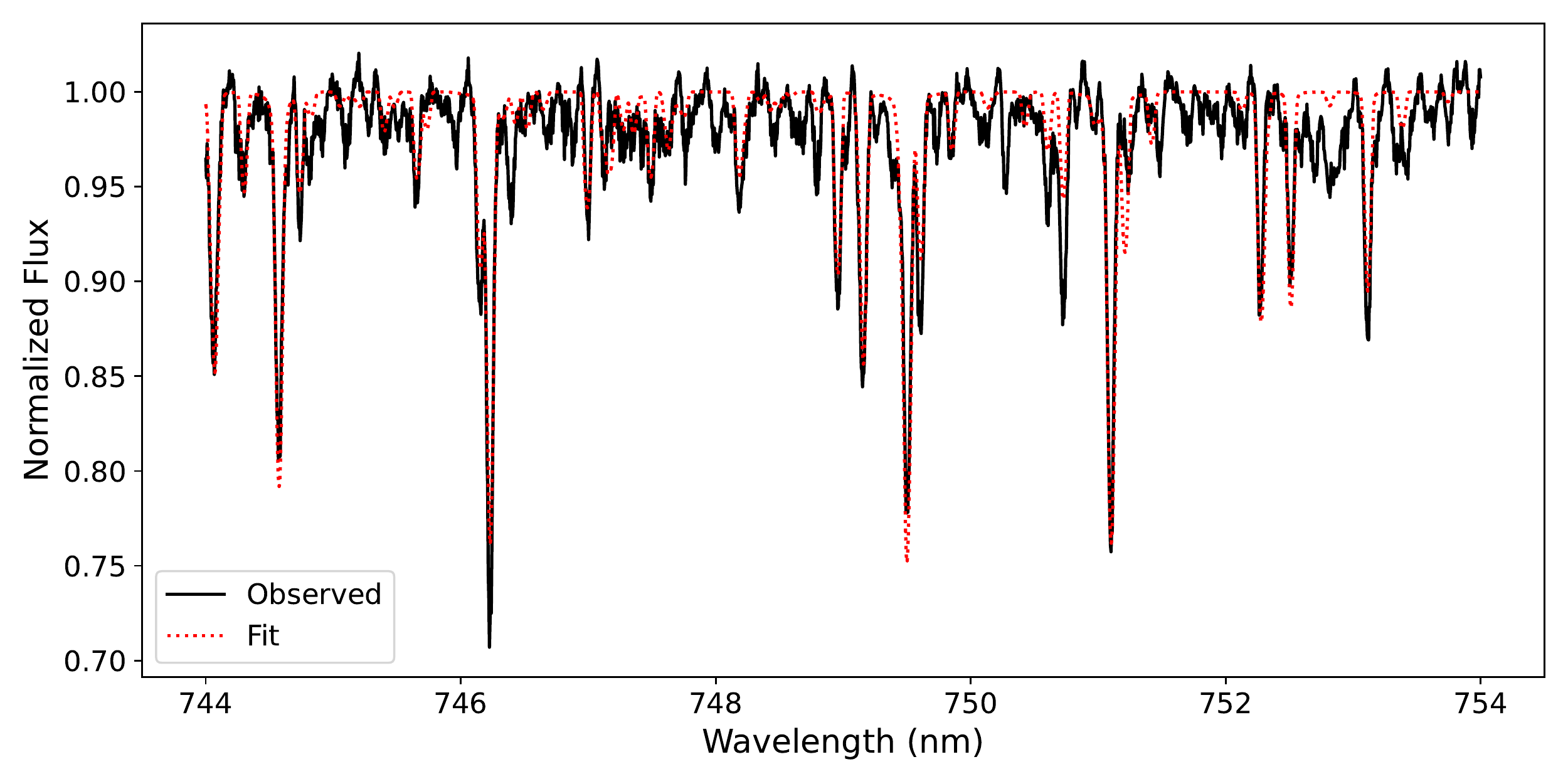}
    \caption{Resulting fit of ZEEMAN synthetic spectra \textit{(red)} of DQ Tau spectrum \text{(black)} over the 744 to 754 nm-range for A (\textit{(top)} and B \textit{(bottom)} components.}
    \label{fig:ZEEMANfit}
\end{figure*}

\section{LSD profile fitting by ZDI procedure}

In this section we provide the results of the LSD Stokes I and V profiles fitting using ZDI.
The profile at $\phi$=0.26 (HJD 2~459~177.9018) has been removed from the analysis because of the low S/N of the Stokes V profile.

\begin{figure*}
    \centering
    \includegraphics[width=.60\textwidth]{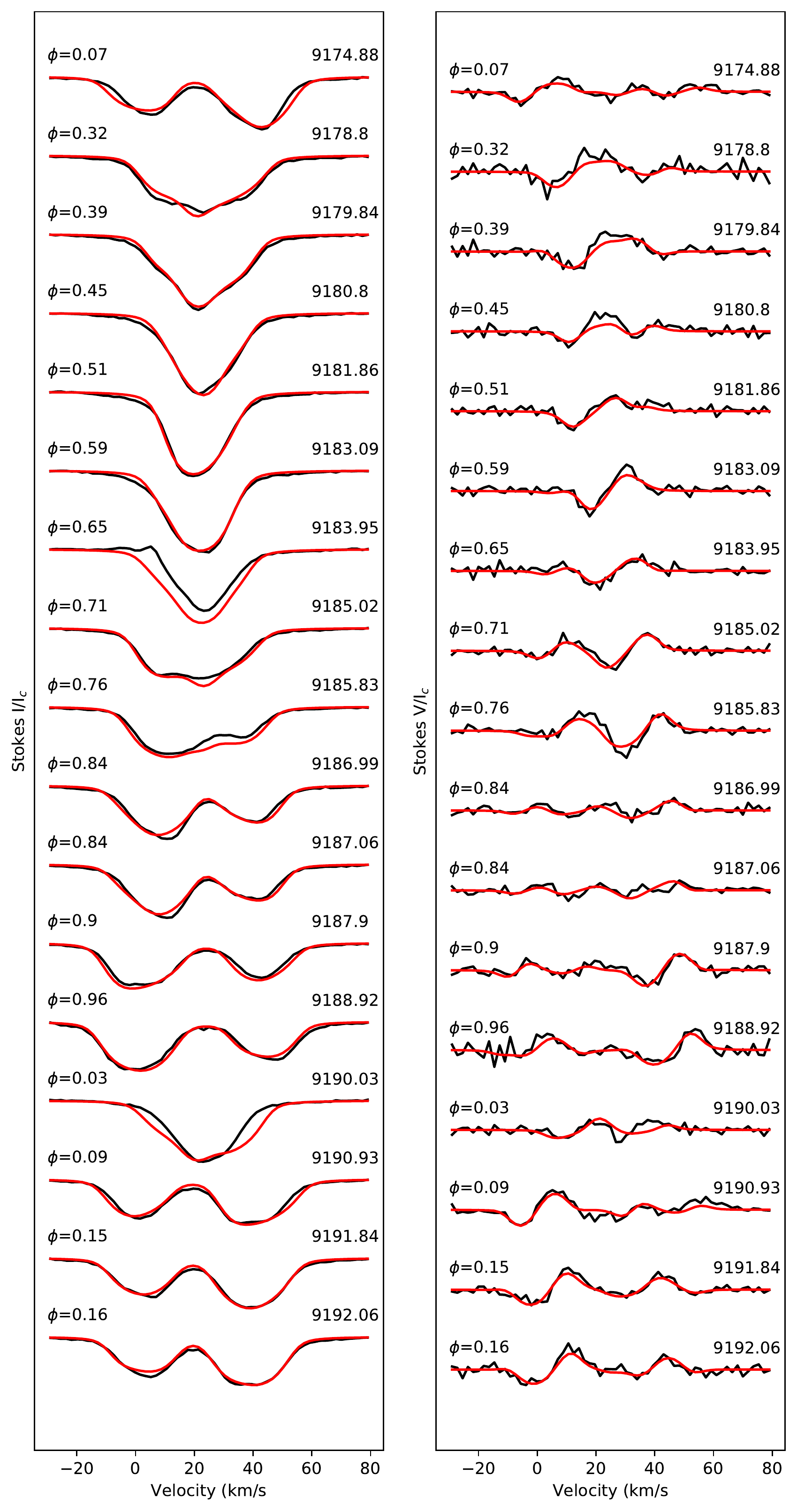}
    \caption{LSD Stokes I \textit{(left)} and V \textit{(right)} profiles analysed with ZDI analysis \textit{(black)} and the resulting model fit \textit{(red)}.}
    \label{fig:ZDIfit}
\end{figure*}

\section{Disentangling procedure of the Ti~{\sc i} used for the Zeeman intensification analyse}

We are presenting in this section the results of the disentangling procedure used in Sec.~\ref{subsec:zeemanInt}. 
This procedure allows to disentangle the spectra of the A, the B, and the telluric components.

\begin{figure*}
    \centering
    \includegraphics[width=.80\textwidth]{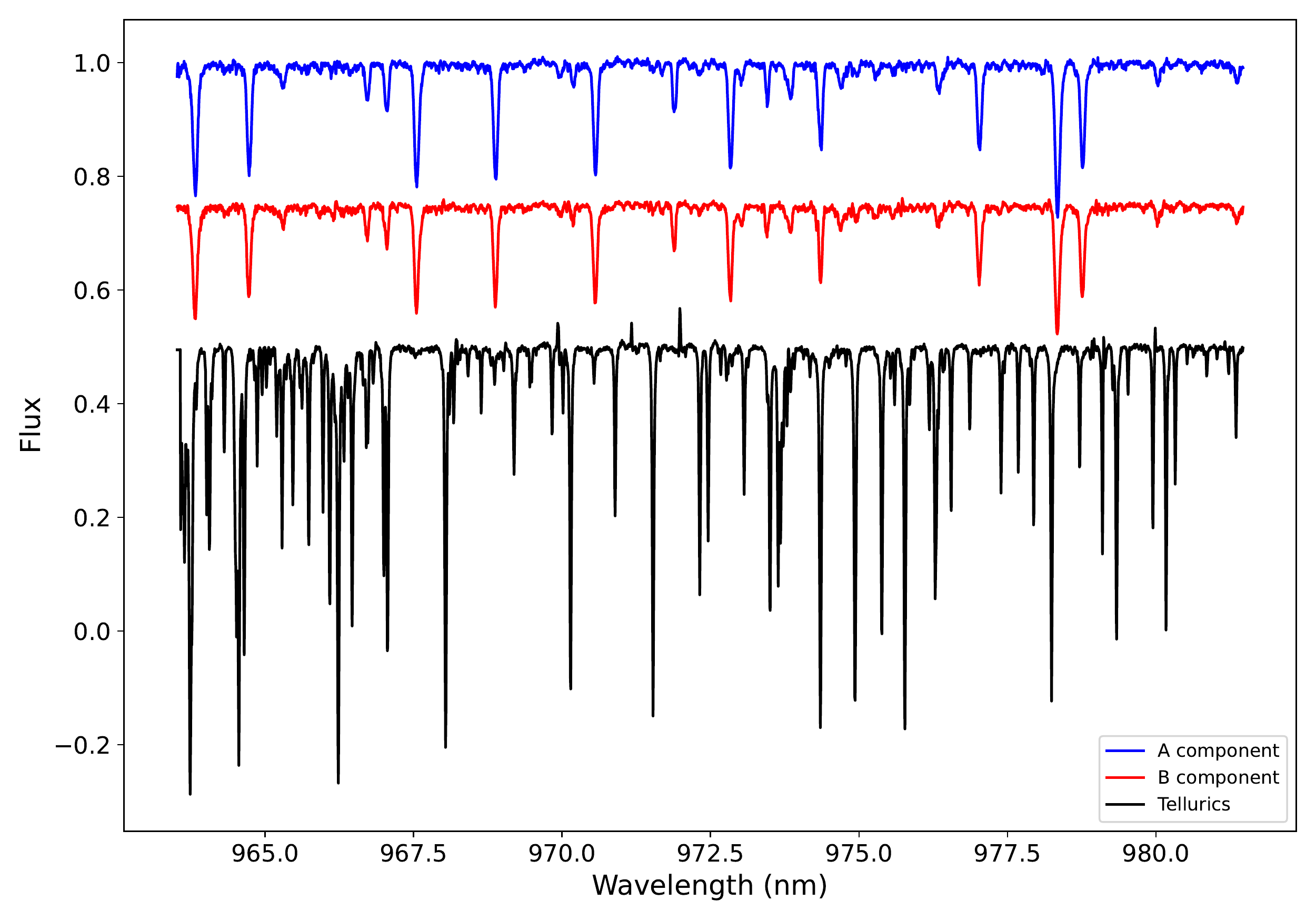}
    \caption{Disentangled spectra of the 963.5 to 981.5 nm region. The spectra of the A \textit{(blue)}, B \textit{(red)}, and tellurics \textit{(black)} components have been shifted vertically for more visibility.}
    \label{fig:disTi}
\end{figure*}

\section{Inference results of Zeeman intensification analysis}

Here we present the corner plot of the posterior distribution for all parameters derived during the Zeeman intensification analysis.
In this analysis, the veiling was not corrected, yielding a luminosity ratio higher than unity.

\begin{figure*}
    \centering
    \includegraphics[width=.99\textwidth]{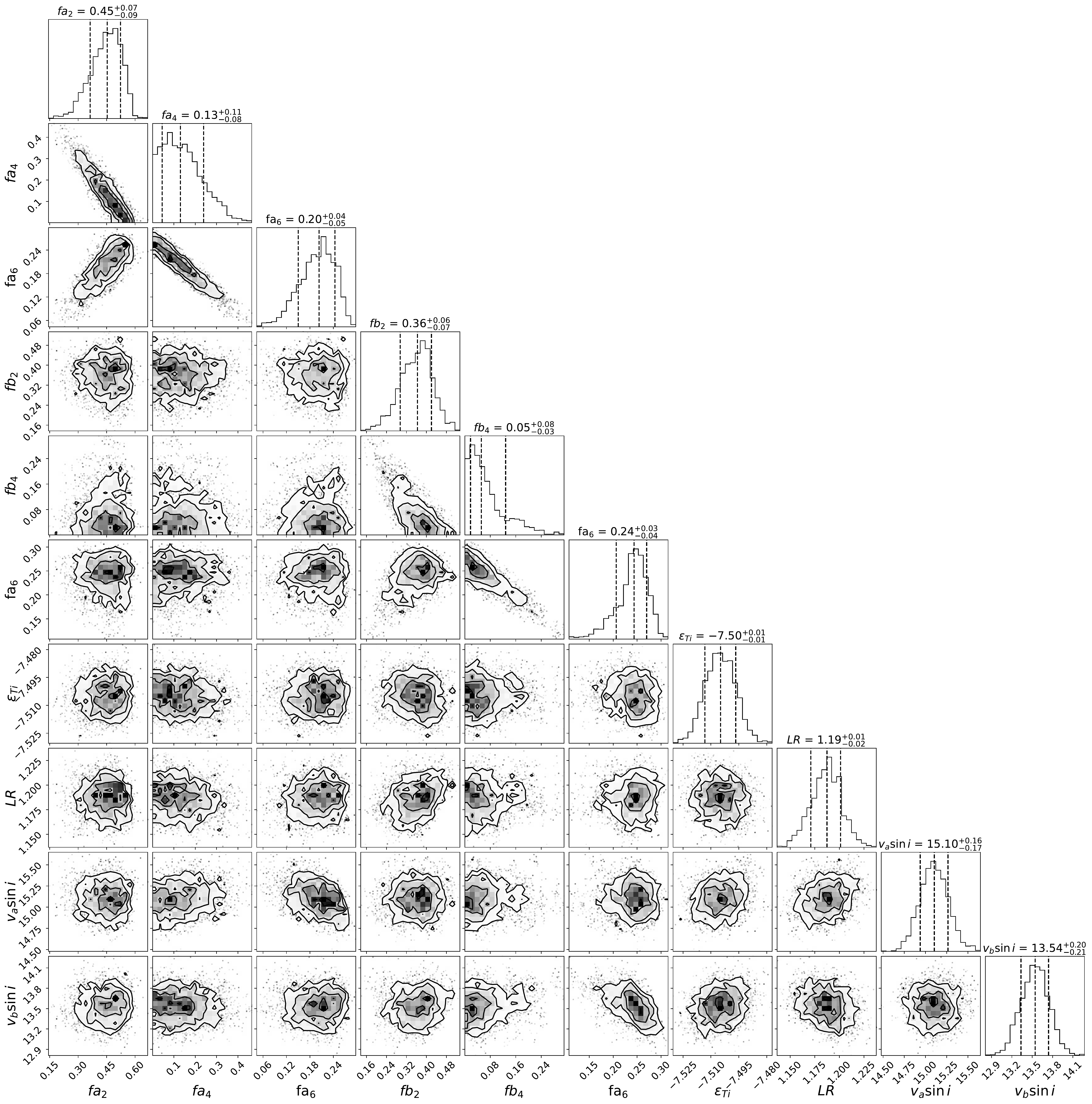}
    \caption{Corner plot showing the posterior distribution of the parameters derived by the Zeeman intensification analysis. \textit{From left to right:} filling factors of the 2~kG, 4~kG and 6~kG magnetic field strengths for A and B components, Ti abundances, luminosity ratio, and $v\sin i$ values for the A and B components}
    \label{fig:corner}
\end{figure*}


\bsp	
\label{lastpage}
\end{document}